\documentclass[pra, twocolumn, amscd, amsmath, amssymb,verbatim,letterpaper,showpacs,superscriptaddress]{revtex4-1}
\usepackage{verbatim}
\usepackage[pdftex]{graphicx}
\usepackage[pdftex]{epsfig}
\newcommand{\bra}[1]{\left\langle{#1}\right\vert}
\newcommand{\ket}[1]{\left\vert{#1}\right\rangle}

\newcounter{subfigure}

\begin{document}
\title{Non-interacting multi-particle quantum random walks applied to the graph isomorphism problem for strongly regular graphs}
\author{Kenneth Rudinger}
\email{rudinger@wisc.edu} 
\author{John King Gamble}
\affiliation{University of Wisconsin-Madison, Physics Department \\ 1150 University Ave, Madison, Wisconsin 53706, USA}
\author{Mark Wellons} 
\author{Eric Bach}
\affiliation{University of Wisconsin-Madison, Computer Sciences Department \\ 1210 W. Dayton St, Madison, Wisconsin 53706, USA}
\author{Mark Friesen}
\author{Robert Joynt}
\author{S. N. Coppersmith}
\email{snc@physics.wisc.edu}
\affiliation{University of Wisconsin-Madison, Physics Department \\ 1150 University Ave, Madison, Wisconsin 53706, USA}

\begin{abstract}
We investigate the quantum dynamics of particles on graphs (``quantum random walks"), with the aim of developing quantum algorithms for determining if two graphs are isomorphic (related to each other by a relabeling of vertices).  We focus on quantum random walks of multiple non-interacting particles on strongly regular graphs (SRGs), a class of graphs with high symmetry that is known to have pairs of graphs that are hard to distinguish.  Previous work has already demonstrated analytically that two-particle non-interacting quantum walks cannot distinguish non-isomorphic SRGs of the same family.  Here, we demonstrate numerically that three-particle non-interacting quantum walks have significant, but not universal, distinguishing power for pairs of SRGs, proving a fundamental difference between the distinguishing power of two-particle and three-particle non-interacting walks.  We analytically show why this distinguishing power is possible, whereas it is forbidden for two-particle non-interacting walks.  Based on sampling of SRGs with up to 64 vertices, we find no difference in the distinguishing power of bosonic and fermionic walks.  In addition, we find that the four-fermion non-interacting walk has greater distinguishing power than the three-particle walks on SRGs, showing that increasing particle number increases distinguishing power.  However, we also analytically show that no non-interacting walk with a fixed number of particles can distinguish all SRGs, thus demonstrating a potential fundamental difference between the distinguishing power of interacting and noninteracting walks.
\end{abstract}

\maketitle

\renewcommand{\thefigure}{A\arabic{subfigure}}
\setcounter{subfigure}{1}

\section{Introduction}
There has long been interest in algorithms that use random walks to solve a variety of mathematical and scientific problems \cite{Motwani1996, Aleliunas1979, Trautt2006, Sessions1997, Kilian2003}.  Typically, the random walks in question have been classical random walks (CRWs).  However, there is increasing interest in random walks with quantum walkers.  In particular settings, these quantum random walks (QRWs) have been shown to have computational advantages over CRWs \cite{Aharonov1993,Bach2004,Solenov2006}.  Certain algorithms utilizing QRWs have been proven to have faster runtimes than their best known classical counterparts \cite{Childs2002, Shenvi2003, Ambainis2003, Ambainis2004, Magniez2007, Potocek2009, Reitzner2009}.\\
\indent Additionally, QRWs have been experimentally demonstrated in a variety of physical settings, such as ion traps \cite{Schmitz2009}, atom traps \cite{Karski2009}, quantum optics \cite{Schreiber2010, Broome2010}, and NMR systems \cite{Ryan2005}.  Recent works have experimentally realized QRWs with two walkers, demonstrating the potential for implementing QRWs with many walkers \cite{Zahringer2010, Owens2011, Peruzzo2010, Sansoni2012}.  Moreover, there are proposed methods for physically implementing non-trivial walks \cite{Manouchehri2009}, indicating that there may be many QRW algorithms to be developed that would both be physically realizable and computationally powerful.\\
\indent Often the context for QRWs is one in which the walks occur on graphs.  It has been shown that QRWs are universal; any quantum algorithm can be mapped onto a QRW on such a graph \cite{Childs2009}.  It is also the case that many interesting computational problems are easily expressed in graph theoretic terms \cite{Gamble2010}.   Thus there is considerable interest in further exploring QRWs on graphs, with the hope that we may be able to use such a framework to solve certain problems.\\
\indent There are also interesting physical phenomena associated with many particles walking on a graph.  It is known that QRWs of non-interacting bosons on graphs can give rise to effective statistical interactions \cite{Burioni2000, Buonsante2002, Mancini2007}.  It has even been shown that Bose-Einstein condensation can occur at finite temperature in less than two dimensions if the bosons are placed on a particular kind of graph \cite{Mancini2007}.  
Therefore, there is motivation in further exploring the dynamics of multi-particle ensembles on graphs.\\
\indent This paper addresses the graph isomorphism problem, which is, given two graphs, to determine if they are isomorphic; that is, if one can be transformed into the other by a relabeling of vertices.  This problem is of note for several reasons.  While many graph pairs may be distinguished by a classical algorithm which runs in a time polynomial in the number of vertices of the graphs, there exist pairs which are computationally difficult to distinguish.  Currently, the best general classical algorithm has a runtime of $O(c^{\sqrt{N} \log N})$, where $c$ is a constant and $N$ is the number of vertices in the two graphs \cite{Spielman1996}.  Graph isomorphism (GI) is believed to be similar to factoring in that both are thought to be NP-Intermediate problems \cite{Schoning1988}.  Additionally, both problems may be approached as hidden subgroup problems, though this approach has had limited success for GI \cite{Moore2007}.  Due to these similarities, and the known quantum speedup available for factoring \cite{Shor1994}, there is hope that there similarly exists a quantum speedup for GI.  \\
\indent Strongly regular graphs (SRGs) are a particular class of graphs that are difficult to distinguish classically \cite{Spielman1996}.  (See Section~\ref{subsec:SRG} for a formal definition.)  Shiau \emph{et al.} showed that the single-particle continuous-time QRW fails to distinguish pairs of SRGs with the same family parameters \cite{Shiau2005}.  Gamble \emph{et al.} extended these results, proving that QRWs of two non-interacting particles will always fail to distinguish pairs of non-isomorphic strongly regular graphs with the same family parameters \cite{Gamble2010}.  They also demonstrated numerically the distinguishing power of the two-boson interacting QRW; it successfully distinguished all tested pairs of SRGs \cite{Gamble2010}.  Since the publication of Gamble \emph{et al.}, Smith proved that for any fixed number of bosons \emph{p}, there exist non-isomorphic graph pairs which the \emph{p}-boson interacting walk fails to distinguish \cite{JSmith2010}.  These counterexample graphs are not strongly regular; whether or not the two-boson interacting walk successfully distinguishes non-isomorphic strongly regular graphs is still an open question.\\
\indent  Investigations into discrete-time QRW algorithms for GI have also been made \cite{Emms2009, Guo2011, Berry2011}.  Berry and Wang numerically showed that a discrete-time non-interacting QRW of two particles could distinguish some SRGs, something its continuous-time counterpart cannot do.  However, this distinguishing power is not universal on SRGs, nor is an analytic explanation of the distinguishing power given \cite{Berry2011}.  The discrete-time algorithm proposed by Emms \emph{et al.} successfully distinguished all tested SRGs \cite{Emms2009}, but it has been shown to not be universal \cite{JSmith2010}; it is unknown if it is universal on SRGs.  Additionally, for the same number of particles, the discrete-time QRWs require Hilbert spaces larger than the ones required by continuous-time QRWs \cite{Berry2011}.
Thus it remains an open question as to whether or not discrete-time walks in general have fundamentally greater distinguishing power than continuous-time walks, or if they are better candidates for a universal GI algorithm.\\
\indent This paper extends the results of \cite{Gamble2010} to address continuous-time multi-particle non-interacting quantum walks on SRGs, with a particular focus on understanding the role of particle number in determining the distinguishing power of the walks.  We have several main results.  We numerically demonstrate that three-particle non-interacting walks have significant (but not universal) distinguishing power on hard-to-distinguish pairs of SRGs.  Additionally, we find that a four-fermion non-interacting walk has even greater (but still not universal) distinguishing power on SRG pairs.  We analytically explain where this distinguishing power comes from, and how these multi-particle non-interacting walks are fundamentally different from single-particle and two-particle non-interacting walks.  This is done by showing that a particular feature present in the smaller walks which limits their distinguishing power is not present in walks of three or more non-interacting particles.  Further, we analytically show that, even though the distinguishing power of non-interacting walks increases with particle number, there is no non-interacting walk with a fixed number of particles that can, with our comparison algorithm, distinguish all strongly regular graphs.\\
\indent This paper is organized as follows.  Section~\ref{sec:Back} covers the requisite background, including graph theoretic definitions and concepts, a review of strongly regular graphs, and a formal definition of the quantum random walk.  In Section~\ref{sec:SRGs}, we first demonstrate analytically how two-particle non-interacting walks are fundamentally different from three-particle non-interacting walks.  We then present the numerical results for non-interacting three-particle and four-particle walks on SRGs.  In the final part of Section~\ref{sec:SRGs}, we demonstrate that a $p$-particle non-interacting QRW cannot distinguish all SRGs for any fixed $p$.
We discuss our conclusions in Section~\ref{sec:Disc}.\\
\indent 
Appendix A discusses a fundamental difference between non-interacting walks of two particles and non-interacting walks of more than two particles.  Appendix B provides details necessary to show that a non-interacting $p$-particle walk cannot distinguish all SRGs for a fixed $p$.  In Appendix C, we show that the number of unique evolution operator elements for a $p$-particle non-interacting walk is super-exponential in $p$.  Lastly, we explain in Appendix D how we ensure numerical stability and determine numerical error in our simulations.
\section{Background}
\label{sec:Back}
\subsection{Basic Graph Definitions}
Here we develop the background and definitions necessary to discuss multi-particle QRWs on graphs.  This paper only considers simple, undirected graphs.  A graph $G = (V, E)$ is a set of vertices $V$ and edges $E$.  The vertices are a set of labels, usually integers, and the edges are a list of unordered pairs of vertices.  If a pair of vertices appears in $E$, then the vertices are connected by an edge; otherwise there is no edge between the vertices and they are considered disconnected.  The terms ``adjacent'', ``neighboring'', and ``connected'' may be used interchangeably to refer to a vertex pair which shares an edge.  It is convenient to represent a graph by its adjacency matrix ${\mathbf A}$, defined as:
\begin{equation}
A_{ij} = \begin{cases} 1 & \text{if vertices $i$ and $j$ are connected.} \\ 0 & \text{if vertices $i$ and $j$ are disconnected.} \end{cases}
\end{equation}
A graph of $N$ vertices has an $N \times N$ adjacency matrix.  For the undirected and simple graphs considered here, ${\bf A}$ is symmetric, with zeros on the diagonal.  \\
\indent 
Two graphs are isomorphic if one graph is transformed into the other by a relabeling of vertices.  More formally, given two adjacency matrices ${\bf A}$ and ${\bf B}$, the graphs represented by ${\bf A}$ and ${\bf B}$ are isomorphic if and only if a permutation matrix ${\bf P}$ exists such that ${\bf B}={\bf P}^{-1}{\bf AP}$.
\subsection{Strongly Regular Graphs }
\label{subsec:SRG}
This paper addresses strongly regular graphs (SRGs), which we examine because they are difficult to distinguish classically, and because of their simple algebraic properties \cite{Spielman1996,Godsil2001}.  An SRG is characterized by four parameters, denoted $(N, k, \lambda, \mu)$.  $N$ is the number of vertices in the graph, and each vertex is connected to $k$ other vertices (the graph is $k$-regular, or has degree $k$).  Each pair of neighboring vertices shares $\lambda$ common neighbors, while each pair of non-adjacent vertices shares $\mu$ common neighbors.  The set of SRGs sharing the same set of four parameters is referred to as an SRG \emph{family}; correspondingly, the four parameters are often called the family parameters.  While some SRG families may have only one non-isomorphic member, there are many families of SRGs with multiple non-isomorphic graphs.  These are the families which are of interest to us.\\
\indent 
The adjacency matrix of any SRG has at most three eigenvalues.  As these eigenvalues and their multiplicities are functions of the family parameters, the adjacency matrices of SRGs in the same family are always cospectral \cite{Godsil2001}.  This contributes to the difficulty of distinguishing non-isomorphic SRGs.\\
\indent 
The adjacency matrix of any SRG satisfies the particularly useful algebraic identity \cite{Godsil2001}:
\begin{equation}
\bold{A}^2 = (k-\mu)\bold{I} + \mu \bold{J} + (\lambda-\mu) \bold{A},
\end{equation}
where $\bold{I}$ is the identity and $\bold{J}$ is the matrix of all ones.  Because $\bold{J}^2=N\bold{J}$, $\bold{JA}=\bold{AJ}=k\bold{A}$, and $\bold{I}$ acts trivially on $\bold{I}$, $\bold{J}$, and $\bold{A}$, we see that $\{\bold{I},\bold{J},\bold{A}\}$ forms a commutative three-dimensional algebra, so we conclude that for any positive integer $n$:
\begin{equation}
\label{eq:SRG}
\bold{A}^n=\alpha_n \bold{I} + \beta_n \bold{J} + \gamma_n \bold{A},
\end{equation}
where $\alpha_n$, $\beta_n$, and $\gamma_n$ depend only on $n$ and the family parameters.
\subsection{Defining the quantum random walk}
Now we discuss how we form a continuous-time non-interacting quantum random walk on a graph.  As in \cite{Gamble2010}, we use the Hubbard model, where each site corresponds to a graph vertex.  A particle can move from one vertex to another if the two vertices are connected.  Thus, for a graph on $N$ vertices with adjacency matrix $\bold{A}$, our non-interacting Hamiltonian is given by
\begin{equation}
\label{eq:HamDef}
\bold{H} = -\sum_{i,j}^N A_{ij} c_i^\dagger c_j,
\end{equation}
where $c_i^\dagger$ and $c_i$ are the creation and annihilation operators, respectively, for a boson or (spinless) fermion at site $i$. For bosons, they satisfy the commutation relations $[c_i,c_j^\dagger]=\delta_{ij}$ and $[c_i,c_j]=[c_i^\dagger,c_j^\dagger]=0$.  For fermions, they satisfy the anti-commutation relations $\{c_i,c_j^\dagger\}=\delta_{ij}$ and $\{c_i,c_j\} = \{c_i^\dagger,c_j^\dagger\}=0$.\\
\indent 
For walks of $p$ bosons, we use basis states of the form $\ket{j_1 \hdots j_p}_B$, which is the appropriately symmetrized basis state with bosons on vertices $j_1$ through $j_p$.  These vertices need not be distinct, since vertices may be multiply occupied.  Similarly, for walks of $p$ fermions, we use basis states of the form $\ket{j_1 \hdots j_p}_F$, which is the appropriately anti-symmetrized basis state with fermions on vertices $j_1$ through $j_p$.  These vertices must be distinct, because the Pauli exclusion principle implies that no vertex can be occupied by multiple fermions.  We refer to these bases as the particles-on-vertices bases.

Following \cite{Gamble2010} and \cite{JSmith2010}, it is straightforward to show that the elements of the $p$-boson or $p$-fermion non-interacting Hamiltonian ($\bold{H}_{p,B}$ and $\bold{H}_{p,F}$, respectively) are, in their respective particles-on-vertices bases:
\begin{align}
_B{\bra{i_1 \hdots i_p}} \bold{H}_{p,B} \ket{j_1 \hdots j_p}_B =\\ -_B{\bra{i_1 \hdots i_p}}\bold{A}^{\oplus p} \ket{j_1 \hdots j_p}_B, \nonumber
\end{align}
\begin{align}
_F{\bra{i_1 \hdots i_p}} \bold{H}_{p,F} \ket{j_1 \hdots j_p}_F =\\ _F{\bra{i_1 \hdots i_p}}\bold{A}^{\oplus p} \ket{j_1 \hdots j_p}_F, \nonumber
\end{align}
where
\begin{align}
\bold{A}^{\oplus{p}}  &= \underbrace{\bold{A} \otimes \bold{I} \otimes \bold{I} \hdots \otimes \bold{I}}_p \\
				&+ \bold{I} \otimes \bold{A} \otimes \bold{I} \hdots \otimes \bold{I} + \hdots + \bold{I} \otimes \bold{I} \otimes \bold{I} \hdots \otimes \bold{A}. \nonumber
\end{align}

The evolution operator is defined in the standard manner:
\begin{equation}
\label{eq:U_first}
\bold{U}(t) = e^{-it\bold{H}},
\end{equation}
where $\hbar=1$ for convenience.
\subsection{Comparison algorithm}
\label{subsec:CompAlg}
Our method for comparing two graphs in an attempt to determine if they are isomorphic or not is the same as the one used in \cite{Gamble2010}.  Given two graphs with adjacency matrices $\bold{A}$ and $\bold{B}$, we compute in the particles-on-vertices basis $\bold{U_A}(t)$ and $\bold{U_B}(t)$, respectively, for the same number and type of particle, as well as the same time $t$.  The absolute value of each element of $\bold{U_A}(t)$ and $\bold{U_B}(t)$ are written to lists $\bold{X_A}$ and $\bold{X_B}$, respectively.  Both lists are sorted, and we compute the distance between the lists, $\Delta$:  
\begin{equation}
\label{eq:Delta}
\Delta = \sum_\nu |\bold{X_A}[\nu] - \bold{X_B}[\nu]|.
\end{equation}
We say that $\bold{A}$ and $\bold{B}$ are distinguished by a particular walk if and only if that walk yields $\Delta \neq 0$; isomorphic graphs and non-isomorphic non-distinguished graphs both yield $\Delta=0$ \cite{Gamble2010}.  We note that we lose phase information by taking the absolute value of the elements, but it makes our comparison procedure more tractable, and seems to do no harm, see \cite{Gamble2010}.  Lastly, for all simulations presented in this paper, $t=1$.

\section{QUANTUM RANDOM WALKS ON STRONGLY REGULAR GRAPHS}
\label{sec:SRGs}
\subsection{Comparing distinguishing power of two- and three-particle non-interacting walks}
\label{subsec:2v3}

In this subsection we show analytically that there is a fundamental difference between two-particle non-interacting walks and three-particle non-interacting walks on strongly regular graphs, because three-particle non-interacting walks are capable of distinguishing SRGs from the same family, unlike two-particle non-interacting walks.  To show this difference, we recall the proof used by Gamble \emph{et al.} to demonstrate the inadequacy of two-particle walks \cite{Gamble2010}.\\
\indent 
The proof in Gamble \emph{et al.} first shows that the value of every element in the two-particle evolution operators ($_B\bra{ij}{\bf U}_{2B}(t)\ket{kl}_B$ or $_F\bra{ij}{\bf U}_{2F}(t)\ket{kl}_F$) must be a function only of the SRG family parameters and the time $t$.  Then it is shown that the multiplicity of each element value in the evolution operator is also a function of SRG family parameters.  We begin similarly here for the three-particle walk, and find that while the values of the elements are all functions of the SRG family parameters, the multiplicities of the values are not.  

We first address the element values.  We refer to each element of each evolution operator (computed in the particles-on-vertices basis) as a \emph{Green's function}, following the nomenclature of Gamble \emph{et al.} \cite{Gamble2010}.  Because the three-particle walk in question is non-interacting, we know that the evolution operator for the walk factorizes into three single-particle evolution operators:
\begin{equation}
\label{eq:U_third-bosons}
_B\bra{ijk}\bold{U}_{3B}\ket{lmn}_B=_B\bra{ijk}{\bold{U}_{1P}}^{\otimes 3}\ket{lmn}_B,
\end{equation}
\begin{equation}
\label{eq:U_third-fermions}
_F\bra{ijk}\bold{U}_{3F}\ket{lmn}_F=_F\bra{ijk}\overline{\bold{U}_{1P}}^{\otimes 3}\ket{lmn}_F,
\end{equation}
where ${{\bf U}_{1P}} ^{\otimes 3} = {\bf U}_{1P} \otimes {\bf U}_{1P} \otimes {\bf U}_{1P}$; ${\bf U}_{1P}$ is the evolution operator for the single-particle walk, that is, ${\bf U}_{1P} = e^{i{\bf A}t}$ and $\overline{\bf U_{1P}} = e^{-i{\bf A}t}$.

Recalling Eq.~\eqref{eq:SRG}, and expanding $e^{i{\bf A}t}$ as a Taylor series in powers of $\bold{A}t$, we note that:
\begin{equation}
\label{eq:U1P_SRG}
\bold{U}_{1P}=\alpha \bold{I} + \beta \bold{J} + \gamma \bold{A},
\end{equation}
where $\alpha$, $\beta$, and $\gamma$ are functions of the family parameters and the time $t$.
Therefore, we conclude that all possible values of the elements of $\bold{U}_{3B}$ and $\bold{U}_{3F}$ (the Green's functions) are determined by the family parameters.  Thus, the set of all potential values for the Green's functions are the same for any two graphs in the same family.  Any distinguishing power of the walks must come from the existence of at least one Green's function with different \emph{multiplicities} for non-isomorphic graphs in the same family.

Gamble \emph{et al.}, prove that the multiplicity of each Green's function for two-particle non-interacting walks is a function of the SRG family parameters.  In Appendix A, we show that there exist Green's functions for the three-particle non-interacting walk on SRGs whose multiplicities are \emph{not} functions of the family parameters.  This is because the multiplicity of a Green's function in a $p$-particle walk depends on how many shared neighbors a collection of up to $p$ vertices has.  For $p=2$, strong regularity uniquely determines the number of shared neighbors: $\lambda$ if the vertices are connected, and $\mu$ if they are not.  However, for $p\geq3$, the multiplicity is dependent on the number of shared neighbors among sets of $p$ vertices.  Thus the multiplicity is not uniquely determined by strong regularity, so the multiplicity for such a Green's function need not be a function of the family parameters.

The definition of SRGs does not directly constrain the number of neighbors of a set of $p$ vertices with $p\geq3$.  However, this difference from the two-particle case does not guarantee that walks of three or more particles can distinguish non-isomorphic SRGs, only that they have the potential to do so.  Our numerical investigations of the distinguishing power of these walks are presented in Section~\ref{subsec:Results}.

\subsection{Numerical results}
\label{subsec:Results}

\begin{table*}[tbh]
\caption{Numerical results for the three-particle non-interacting walks on twelve families of SRGs.  The first column lists the family parameters for the particular SRG family being examined.  The second column lists the number of graphs in the family that we compared.  This number is equal to the number of graphs in the family, with the exception of (49,18,7,6), where we only examined a subset of the family.  The third column gives the number of comparisons made for each family, which is equal to the number of graphs in that family that we examined choose 2.  The fourth and fifth columns list the number of graph pairs which the three-boson and three-fermion walks fail to distinguish, respectively.  We see that out of $70\,712$ graph comparisons, both the boson and fermion walks fail a total of 256 times, corresponding to a success rate of greater than $99.6\%$ \label{tab:TABLE1}}
 \begin{tabular} {| c | c | c | c | c |}
 \hline
 SRG Family ($N$, k, $\lambda$, $\mu$) & Number of Graphs & Comparisons & Boson Failures & Fermion Failures\\
\hline
 (16, 6, 2, 2) & 2 & 1 & 0 & 0\\
\hline
 (16, 9, 4, 6) & 2 & 1 & 0 & 0\\
\hline
 (25, 12, 5, 6) & 15 & 105 & 0 & 0\\
\hline
 (26, 10, 3, 4) & 10 & 45 & 1 & 1\\
\hline
 (28, 12, 6, 4) & 4 & 6 & 0 & 0\\
\hline
 (29, 14, 6, 7) & 41 & 820 & 0 & 0\\
\hline
 (35, 18, 9, 9) & 227 & 25651 & 38 & 38\\
\hline
 (36, 14, 4, 6) & 180 & 16110 & 89 & 89\\
\hline
 (40, 12, 2, 4) & 28 & 378 & 8 & 8\\
\hline
 (45, 12, 3, 3) & 78 & 3003 & 7 & 7\\
\hline
 (49, 18, 7, 6) & 147 & 10731 & 21 & 21\\
\hline
 (64, 18, 2, 6) & 167 & 13861 & 92 & 92\\
\hline
\end{tabular}
\end{table*}

\begin{table*}[tbh]
\caption{Numerical results for four-fermion non-interacting walks on 136 graph pairs that are not distinguished by three-particle non-interacting walks.  Of the 136 graph pairs tested, only one pair is not successfully distinguished.  We therefore see that increasing the number of non-interacting particles beyond three continues to increase the distinguishing power of the non-interacting QRWs.\label{tab:TABLE2}}
\begin{tabular} {| c | c | c |}
\hline
Family ($N$, $k$, $\lambda$, $\mu$) & 3 Particle Failures & 4 Fermion Failures \\
\hline
(26, 10, 3, 4) & 1 & 0 \\
\hline
(35, 18, 9, 9) & 38 & 0 \\
\hline
(36, 14, 4, 6) & 89 & 1 \\
\hline
(40, 12, 2, 4) & 8 & 0 \\
\hline
\end{tabular}
\end{table*}
In this subsection, we present our numerical results for three-particle and four-fermion walks on SRGs.  To simulate a walk on a graph, we compute the appropriate Hamiltonian and exponentiate it to compute its corresponding evolution operator, following the algorithm described in Section~\ref{subsec:CompAlg}.  Then, to compare pairs of non-isomorphic graphs from the same family, we compute the list distance $\Delta$, defined in Equation~\eqref{eq:Delta}.  We find our error on $\Delta$ to be no greater than $10^{-6}$, so two non-isomorphic graphs are considered distinguished if and only if $\Delta >10^{-6}$.  Further details of numerical error analysis are provided in Appendix D.\\
\indent Because the Hamiltonians are very large, we must use a sparse matrix exponentiation routine \cite{expokit} to make exponentiation computationally tractable.  (The largest evolution operators we compute have a dimension of $91\,390$, and correspond to the four-fermion walks on graphs of 40 vertices.) Additionally, in order to be able perform these exponentiations sufficiently quickly, we parallelize the computations, utilizing the Open Science Grid and the University of Wisconsin-Madison's Center for High Throughput Computing Cluster.\\
\indent Our numerical results for three-particle walks are presented in Table~\ref{tab:TABLE1}.  For the $70\,712$ pairs of SRGs compared, the boson and fermion walks distinguish all but 256 pairs, corresponding to a success rate of greater than $99.6\%$.  Thus we see that both the three-boson and three-fermion walks have significant (but not universal) distinguishing power on SRGs, while the two-particle non-interacting walks fail on \emph{all} pairs of non-isomorphic graphs in the same family \cite{Gamble2010}.\\
\indent The bosonic and fermionic walks fail to distinguish the same pairs of non-isomorphic graphs that we tested; we have found no graph pair that one kind of particle successfully distinguishes while the other does not.  Thus, despite having a state space of smaller dimension (due to Pauli exclusion), the three-fermion walk has the same distinguishing power as the three-boson walk on all tested graph pairs.  It remains an open question whether graph pairs exist for which this is not true.\\
\indent Having identified some graph pairs that three non-interacting particles fail to distinguish, we want to know if non-interacting walks exist that can distinguish these graphs.  However, it is computationally expensive (even with speedup provided by parallelization) to simulate four-particle walks.  We therefore simulated only fermion walks, and only on a subset of the three-particle counterexample graph pairs.  Our results are summarized in Table~\ref{tab:TABLE2}.  We simulated four-fermion non-interacting walks on 136 counterexample pairs, finding that all but one pair are distinguished.\\
\indent Since increasing the number of non-interacting particles in the walk apparently increases the distinguishing power, it is natural to ask ``Does there exist a $p$ such that the $p$-particle non-interacting walk can distinguish all strongly regular graphs?"  The next subsection shows that the answer to this question is no.
\subsection{Limitations of non-interacting walks}
\label{subsec:R}
In this subsection, we show that pairs of non-isomorphic strongly regular graphs exist that are not distinguished by any $p$-particle non-interacting quantum walk with fixed $p$ in conjunction with the comparison algorithm described by Equation~\eqref{eq:Delta}.  This is because for a fixed $p$, there exists an $N$ such that the number of strongly regular graphs with $N$ vertices is larger than the maximum number of graphs distinguishable by the $p$-particle non-interacting walk.

To prove this claim, we define $S(N)$, the number of strongly regular graphs in a particular family with $N$ vertices, and $Z(p,N)$, the number of distinct ``graph fingerprints" that the $p$-boson walk can generate for an SRG family whose graphs have $N$ vertices.  By a ``graph fingerprint,'' we mean a sorted list of the absolute value of every element of an evolution operator (Eq.~\eqref{eq:U_third-bosons}).  We examine the boson walk here, because its state space is strictly larger than the fermion walk of the same number of particles.  Thus the $p$-boson walk generates more fingerprints than the $p$-fermion walk (even though we have seen no evidence yet that it distinguishes more graph pairs).  Therefore, $Z(p,N)$ bounds from above the maximum number of SRGs with $N$ vertices in a particular family that non-interacting walks of either $p$ fermions or bosons can distinguish.

We now define the ratio $R(p,N)$:
\begin{equation}
R(p,N)=\frac{S(N)}{Z(p,N)}.
\end{equation}
We will show that for any fixed $p$, $R$ is greater than 1 for large enough $N$, thus demonstrating that there exist more SRGs than the $p$-particle walk can distinguish. 

It is shown in \cite{Stones2010} that there is a mapping between Latin squares of size $n$ and SRGs of size $n^2$ with family parameters $(n^2, 3(n-1), n,6)$.  The results of \cite{McKay2005,Stones2010}, imply that when $N$ is large enough, the number of non-isomorphic Latin square SRGs of size $N$ is bounded below by:
\begin{equation}
S(N) \geq \frac{1}{6} (\sqrt{N}!)^{2\sqrt N - 3} N^{\tfrac{-N}{2}}.
\end{equation}

As for $Z(p,N)$, we show in Appendix B that for a fixed $p$, $Z$ satisfies the inequality:
\begin{equation}
Z(p,N) < N^{2X_p (p+1)},
\end{equation}
where $X_p$ is the number of unique values a Green's function for a $p$-boson walk can assume.  While it can be shown that $X_p$ is super-exponential in $p$, it does not depend on $N$.  This is because the value of a Green's function for a non-interacting $p$-particle QRW on an SRG is determined by a configuration of up to $2p$ vertices in that SRG, as discussed in Section~\ref{subsec:2v3} and Appendix A.

To examine the behavior of $R$ in the limit of large $N$, we use Stirling's formula:
\begin{equation}
x!= \sqrt{2 \pi} e^{-x} x^{x+1/2}(1+O(x^{-1})).
\end{equation}
This allows us compute a lower-bound for $R$ in the limit of large $N$:
\begin{equation}
\label{eq:R_lim}
\displaystyle 	\lim_{N\to\infty} R \geq \dfrac{1}{6} (2 \pi)^{\sqrt N - \tfrac{3}{2}} e ^ {-2 N + 3 \sqrt{N}} N^{\tfrac{N}{2} - \sqrt N - \tfrac{3}{4} -2X_p (p+1)}.
\end{equation}
Taking the logarithm of Eq.~\eqref{eq:R_lim} yields:
\begin{equation}
\displaystyle 	\lim_{N\to\infty} \log R(p,N) \geq  \displaystyle 	\lim_{N\to\infty}{\frac{N}{2}\log N}+O(N),
\end{equation}
which diverges as $N\rightarrow \infty$.  Therefore, for a fixed $p$, $R$ approaches $\infty$ as $N$ increases, showing that no $p$-particle non-interacting walk can distinguish all SRGs.

One can let $p$ grow slightly with $N$ and achieve the same result.  Indeed, we show in Appendix C that 
\begin{equation}
\label{eq:log2Xp}
\log_2(X_p) = p^2 + O(p \log p).
\end{equation}
Using this, we find our argument remains valid for \\ $p < C \sqrt{\log_2 N}$, for any $C < 1$.

We can contrast these results to those of Gamble \emph{et al.}.  They found that the hard-core two-boson walk distinguished all graph pairs in a dataset of over 500 million pairs of SRGs \cite{Gamble2010}.  It is an open question as to whether or not the two-boson hard-core walk has universal distinguishing power on SRGs.  Even if does not, it is still possible that there exists a fixed $p>2$ such that the $p$-boson hard-core walk could distinguish all SRGs.  If this is the case, then this would be a marked difference between the non-interacting and hard-core walks.
\section{Discussion}
\label{sec:Disc}
We have shown how three-particle non-interacting quantum random walks are qualitatively different from two-particle non-interacting quantum random walks; the latter will always fail to distinguish non-isomorphic strongly regular graphs from the same family, whereas the former successfully distinguish many (but not all) non-isomorphic pairs of strongly regular graphs.  We have analytically identified a fundamental difference between these two classes of quantum walks.  The three-particle walks have potential distinguishing power because the shared connectivity of triples of vertices in SRGs is not governed by the SRG family parameters.  We have also demonstrated numerically that three-particle non-interacting walks have significant, but not universal, distinguishing power on SRGs.  We observe numerically that bosonic and fermionic walks distinguish the same pairs of non-isomorphic pairs of graphs.  Increasing the number of non-interacting fermions to four further increases distinguishing power.  However, this distinguishing power is not limitless; we have shown that for any fixed number of non-interacting particles, there exist non-isomorphic pairs of SRGs that cannot be distinguished.

Lastly, we discuss the implications of these results in terms of the computational complexity of the graph isomorphism problem.  Not only are there graph pairs on which the three- and four-particle walks fail, but we know that for any fixed particle number, there will be SRGs that such non-interacting walks cannot distinguish.  It is still possible that, given any non-isomorphic SRG pair of a fixed size $N$, there exists a $p$ such that the $p$-particle non-interacting walk will succeed in distinguishing the graphs.  However, the lower bound given at the end of Section~\ref{subsec:R} rules out the possibility of our algorithm providing a classical polynomial-time solution to GI for SRGs.

\section{Acknowledgements}
This work was supported in part by ARO, DOD (W911NF-09-1-0439), NSF (CCR-0635355), and the National Science Foundation Graduate Research Fellowship under Grant No. (DGE-0718123).  We thank Dong Zhou, Dan Bradley, and Alessandro Fedrizzi for useful discussions.  We also thank the HEP, Condor, and CHTC groups at UW-Madison for all their assistance with the numerical computations.  KR thanks Paul Hinrichs for computational assistance.
\section{Appendices}

\subsection{Computing multiplicities of values of matrix elements of the evolution operator for strongly regular graphs}
\label{app:1}
\begin{figure}
\includegraphics{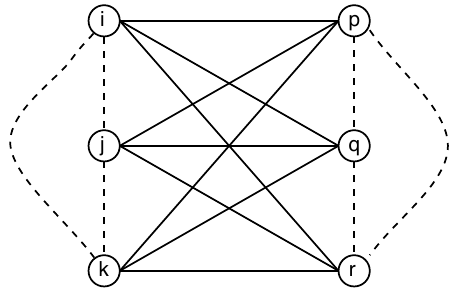}
\caption{\label{fig:CompleteWidget-3}Sketch of a generalized subgraph, or ``widget,'' used to calculate the values and degeneracy of a Green's function for a three-particle quantum walk on an SRG.  The vertices on the right side correspond to the vertices the particles are on to begin with (the ket $\ket{pqr}_B$ or $\ket{pqr}_F$), and the vertices on the left side correspond to the vertices the particles end up on (the bra $_B\bra{ijk}$ or $_F\bra{ijk}$), after application of the evolution operator $U$.  A solid line between vertices $x$ and $y$ indicate that $A_{xy}=1$.  A dashed line between $x$ and $y$ means that the value of $A_{xy}$ does not affect the value of the Green's function.  Thus, for bosons, the depicted widget corresponds to the Green's function $_B\bra{ijk}{\bf U}_{3B}\ket{pqr}_B$ when all six vertices are distinct, and when $A_{xy}=1$ for all $x\in\{i,j,k\}$ and $y\in\{p,q,r\}$.  Eqs.~\eqref{eq:U_third-bosons} and~\eqref{eq:U1P_SRG} show that the value of this Green's function, or widget, is $_B\bra{ijk}{\bf U}_{3B}\ket{pqr}_B=6(\beta+\gamma)^3$.}
\end{figure}
\addtocounter{subfigure}{1}
\begin{figure}
\includegraphics{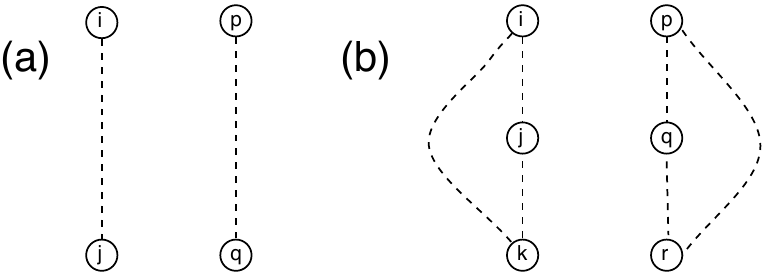}
\caption{\label{fig:widgets2and3}Empty widgets for two-particle and three-particle non-interacting walks.  In both widgets, all vertices are distinct and no vertex in the initial state is adjacent to any vertex in the final state.  The values of the widgets depend only on the family parameters for both (a) and (b), while the degeneracies of these values depend only on family parameters for two particles but not for three.  The multiplicity of each widget's respective Green's function for a particular SRG is equal to the number of times that widget appears in the SRG.  (a)  The empty widget for two particles.  The number of times this widget appears in an SRG is a function of the SRG family parameters, as is the case for all two-particle widgets \cite{Gamble2010}.  (b)  The empty widget for three particles.  The number of times this widget appears in an SRG is \emph{not} a function solely of SRG family parameters.  This is demonstrated by the graphs in Figure~\ref{fig:16}.  An analytic explanation for this phenomenon is given in the text following Equation~\eqref{eq:2-particle-empty}.}
\end{figure}
\addtocounter{subfigure}{1}
\begin{figure}
\includegraphics{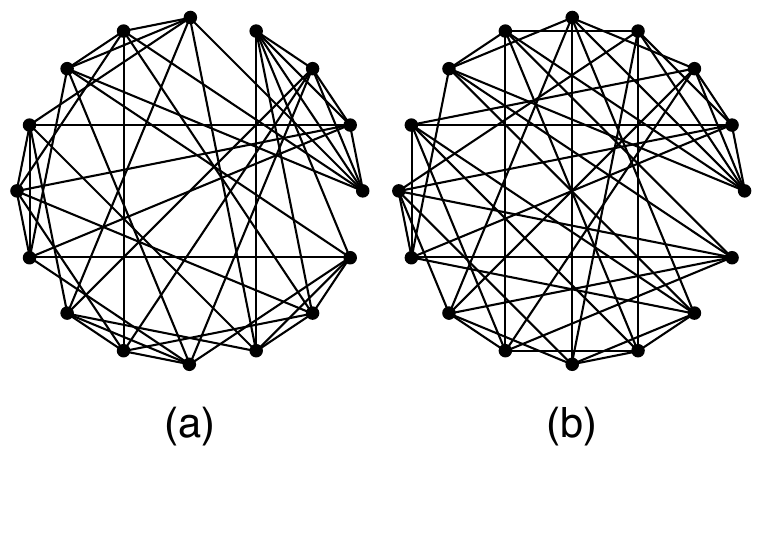}
\caption{\label{fig:16} The two non-isomorphic graphs of the SRG family (16,6,2,2).  The widget of Figure~\ref{fig:widgets2and3}(b) appears in the graph shown in (a) 608 times, whereas the same widget appears in the graph shown in (b) 512 times.  Thus we see that the same three-particle widget can have different multiplicities in graphs of the same family, so the three-particle non-interacting walk can distinguish at least some non-isomorphic graphs from the same SRG family.}
\end{figure}
\addtocounter{subfigure}{1}
\begin{figure}
\includegraphics{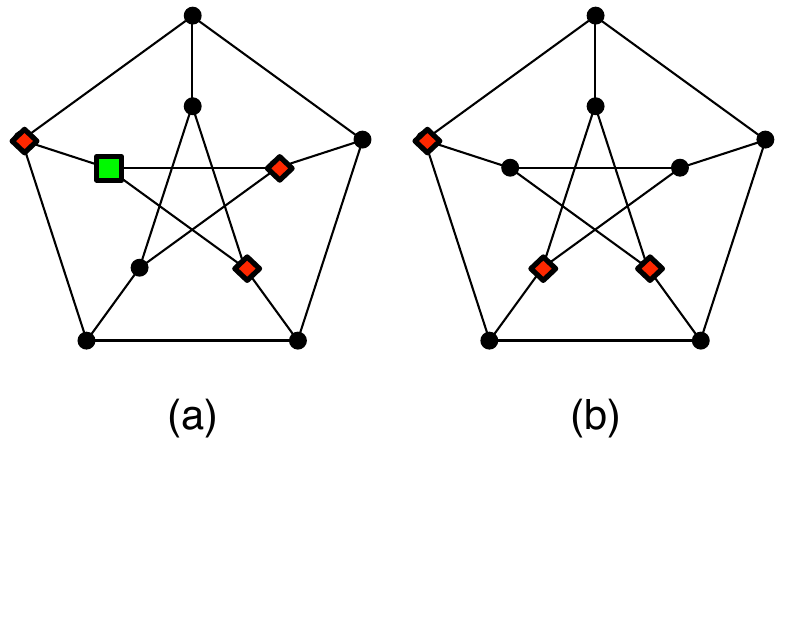}
\caption{\label{fig:Petersens}  (Color online.)  Two copies of the Petersen graph, an SRG with parameters $(10,3,0,1)$.  In each graph, three mutually non-adjacent vertices are highlighted as red diamonds.  In (a), the three vertices share one common neighbor, marked as a green square.  In (b), the three vertices share no common neighbors.  This demonstrates that the number of neighbors common to a triple of vertices in a strongly regular graph is not strictly a function of the SRG family parameters, thus showing why widget multiplicity is not strictly governed by family parameters when $p\geq3$.}
\end{figure}
\addtocounter{subfigure}{1}
Here we discuss how to compute the multiplicities of values of elements of evolution operators, or Green's functions, for SRGs.  We show in this appendix that the multiplicity of a non-interacting three-particle Green's function is in general not a function of SRG family parameters.  This result is used in Section~\ref{subsec:2v3} to demonstrate how two-particle and three-particle non-interacting walks have different distinguishing powers for SRGs.

To compute the multiplicity of each value of the Green's function in a non-interacting three-particle walk, we first note that Eqs.~\eqref{eq:U_third-bosons} through~\eqref{eq:U1P_SRG} imply that the value of a given Green's function depends on the relationships between the vertices in the final state (the bra; $\{i,j,k\}$) and the vertices in the initial state (the ket; $\{p,q,r\}$).  For each pair of indices $(x,y)$, with $x$ from the bra ($x\in \{i,j,k\}$), and $y$ from the ket ($y\in \{p,q,r\}$), there are three possible relations.  The vertices can be connected ($A_{xy}=1$), the vertices can be the same ($\delta_{xy}=1$), or the vertices can be different and disconnected ($A_{xy}=\delta_{xy}=0$).  Therefore, we may think of each Green's function as corresponding to a generalized subgraph of the original graph.  We say ``generalized subgraph" because the Green's function is unaffected by internal connections within the initial or final state; we adopt the more compact terminology of referring to these generalized subgraphs as ``widgets."  

To illustrate this point, let us consider the widget shown in Figure~\ref{fig:CompleteWidget-3}.  The solid lines in the widget indicate that the sites are connected in the graph, while the dashed lines indicate that the value of the widget does not depend on whether or not those sites are connected.  Thus, two widgets are considered the same whether or not sites that are connected by dashed lines are actually adjacent.  To evaluate $_B\bra{ijk} \bold{U}_{3B} \ket{pqr}_B$ for the widget shown in Figure~\ref{fig:CompleteWidget-3}, we note that all six vertices ($\{i,j,k,l,p,q,r\}$) are distinct.  We can then use Eqs.~\eqref{eq:U_third-bosons} and~\eqref{eq:U1P_SRG} to find that $_B\bra{ijk} \bold{U}_{3B} \ket{pqr}_B=6(\beta+\gamma)^3$, where $\beta$ and $\gamma$, defined in~\eqref{eq:U1P_SRG}, are functions of the SRG family parameters.  The multiplicity of this particular value for a particular graph is given by the number of times its corresponding widget occurs in the graph.

To compute the multiplicity, $M$, of $6(\beta+\gamma)^3$ in ${\bf U}_{3B}$, we count the number of occurrences of this widget in the graph.  To do this, we perform the following combinatorial sum, generalizing the procedure outlined in Appendix B of Gamble \emph{et al.} \cite{Gamble2010}.
\begin{align}
\label{eq:widgetcomplete}
M & = \sum_{i<j<k}\sum_{p<q<r} A_{ip} A_{iq} A_{ir} A_{jp} A_{jq} A_{jr} A_{kp} A_{kq} A_{kr}\\
     & = \frac{1}{36}\sum_{ijkpqr} A_{ip} A_{iq} A_{ir} A_{jp} A_{jq} A_{jr} A_{kp} A_{kq} A_{kr} \times \nonumber \\
      & (1-\delta_{ij})(1-\delta_{ik})(1-\delta_{jk})(1-\delta_{pq})(1-\delta_{pr})(1-\delta_{qr}). \nonumber
\end{align}
The analogous sums considered in Gamble \emph{et al.}, which only examines two-particle walks, can be decomposed into sums and traces over powers of the adjacency matrix.  Such operations are given by contracting over two occurrences of the same index in the summand.  Conveniently, these quantities are expressible in terms of SRG family parameters, as is illustrated in Gamble \emph{et al.}  Things are not so simple, however, for the three-particle walks.  By inspection, we see that Eq.~\ref{eq:widgetcomplete} contains contractions over three occurrences of the same index.  Such contractions correspond to neither matrix multiplication nor computing the trace, and cannot in general be massaged into forms expressible in terms of SRG family parameters, as evidenced by the fact that the three-particle walks have distinguishing power over many pairs of SRGs.

However, the above statement does not give us analytic proof that there exist Green's functions whose multiplicities are not functions of the family parameters; up to this point, we are still relying on the numerical results as proof.  Below, we analytically demonstrate that there exist widgets whose multiplicities \emph{cannot} be determined by family parameters.
To demonstrate this, we take a step back to the two-particle walk.  Consider the widget shown in Figure~\ref{fig:widgets2and3}(a).  We can determine this widget's multiplicity for an arbitrary SRG with family parameters $(N,k,\lambda,\mu)$ by performing the combinatorial sum analogous to Equation~\eqref{eq:widgetcomplete}, or equivalently, we can actually count the number of times we can fit this widget on the SRG.  To begin, we pick two sites in the graph to serve as sites $i$ and $j$; these sites may be adjacent or not, as indicated by the dashed line between them in the figure.  Now we must count, given our choice of $i$ and $j$, how many sites we may pick as $p$ and $q$.  If $i$ and $j$ are connected, there are ${N-2k+\lambda} \choose 2$ choices for $p$ and $q$, whereas if $i$ and $j$ are disconnected, there are ${N-2-2k+\mu} \choose 2$ choices for $p$ and $q$.  There are $\frac{Nk}{2}$ choices for connected sites that can serve as $i$ and $j$, and ${N \choose 2} - \frac{Nk}{2}$ disconnected sites.  Thus, the number of four-vertex empty widgets occurring in a two-particle non-interacting walk is:
\begin{align}
\label{eq:2-particle-empty}
M_{2,empty} = \frac{Nk}{2} {N-2k+\lambda \choose 2} + \\ \left( {N \choose 2}-\frac{Nk}{2}\right){N-2-2k+\mu \choose 2}, \nonumber
\end{align}
in agreement with the result in Gamble \emph{et al.} for this particular widget \cite{Gamble2010}.  Thus we see that this widget's multiplicity is, as expected, a function of the family parameters.
Let's see what happens when we try this same approach for the corresponding widget in the three-particle walks, shown in Figure~\ref{fig:widgets2and3}(b).  Again, we consider the multiplicity of the widget in an arbitrary SRG by counting the number of times we can fit this widget on the graph.  Now we pick three sites to serve as $i$, $j$, and $k$.  We want to count, given our choice of $i$, $j$, and $k$, the number of sites that can serve as $p$, $q$, and $r$.  To do this, we need to know how $i$, $j$, and $k$ are connected amongst themselves, just as we did in the previous example.  There are four possible non-isomorphic connectivities, as there can be between zero and three connections amongst these sites.  In order to count the multiplicity of this widget, we must consider for each of these four cases how many sites in the graph are mutually disconnected from sites $i$, $j$, and $k$.  In the previous example, we could answer the analogous question via the family parameters, as illustrated above.  However, this is because the family parameters $\mu$ and $\lambda$ tell us how many common neighbors \emph{pairs} of vertices have.  There are no family parameters which encode this information for \emph{triples} of vertices, as strong regularity does not place absolute constraints on shared connectivities for triples of vertices.  

We illustrate this point with an example in Figure~\ref{fig:Petersens}.  Two copies of the Petersen graph, an SRG with family parameters $(10,3,0,1)$ are depicted.  The first copy highlights three mutually non-adjacent vertices; this particular triple of vertices has one common neighbor.  The second copy also highlights a triple of mutually non-adjacent vertices, but this triple has \emph{no} shared neighbors.  Thus we have demonstrated by example that strong regularity cannot in general uniquely determine the shared connectivity for triples of vertices.

Moreover, we can see that counting the multiplicity of the widget shown in Figure~\ref{fig:widgets2and3}(b) can be used to distinguish two non-isomorphic graphs from the same SRG family.  Figure~\ref{fig:16} shows the two non-isomorphic graphs in the SRG family (16,6,2,2).  The widget in Figure~\ref{fig:widgets2and3}(b) appears 512 times in the first graph and 608 times in the second graph, thus distinguishing them.

We conclude then that there exist three-particle widgets whose multiplicities \emph{cannot} be functions of family parameters.  Thus, the three-particle non-interacting walks are \emph{not} forbidden from distinguishing non-isomorphic SRGs from the same family, unlike the two-particle non-interacting walks.

\subsection{Computing the number of SRG fingerprints}
\label{app:2}

In Section~\ref{subsec:R}, it is shown that quantum walks of $p$ non-interacting particles cannot distinguish all non-isomorphic pairs of strongly regular graphs.  This is done by showing that $Z(p,N)$, the number of graph fingerprints given by the $p$-boson walk on an SRG family with $N$ vertices, is less than the number of non-isomorphic strongly regular graphs with $N$ vertices, in the limit of large $N$.  This subsection presents the calculation of $Z(p,N)$.

To calculate $Z(p,N)$, we note that if there are $X_p$ possible Green's function values for the $p$-boson walk, and $Y$ elements of the evolution operator $U$, then computing the number of unique fingerprints is equivalent to computing the number of ways one can put $Y$ indistinguishable balls in $X_p$ labeled bins, so that \cite{Tucker2004}
\begin{equation}
\label{eq:Z}
Z(p,N) = {{X_p + Y - 1} \choose {X_p - 1}}.
\end{equation}
We recall that $X_p$ is a function of $p$, but not of $N$.  (We may think of $X_p$ as the number of uniquely-valued widgets that appear in the $p$-boson walk.)  However, $Y$, the number of elements in the evolution operator, will depend on both $p$ and $N$, and we henceforth write it as $Y_{p,N}$.  The dimension of the evolution operator is computed by determining how many different ways $p$ bosons can be put on $N$ vertices, which this is the same problem as computing the number of ways to put $p$ indistinguishable balls into $N$ labeled bins.  The number of elements in the evolution operator is just the square of its dimension, so we find that:
\begin{equation}
\label{eq:Y}
Y_{p,N} = {{N+p-1} \choose p}^2.
\end{equation}
\indent Using Equations~\eqref{eq:Z} and~\eqref{eq:Y}, we now compute an upper bound for $Y_{p,N}$ and $Z$.  It can be shown that ${{n+k-1} \choose {k-1}} \leq n^k$ when $n\geq2$ and $k\geq1$.  Thus
\begin{equation}
Y_{p,N} < {{N+p} \choose p} ^2 \leq N^{2(p+1)} 
\end{equation}
and 
\begin{equation}
Z(p,N) \leq \left(Y_{p,N}\right)^{X_p} < N^{2 X_p (p+1)}.
\end{equation}
Therefore, the maximum number of unique graphs the $p$-boson walk can distinguish is bounded above by $N^{2X_p (p+1)}$.  We use this result in Section~\ref{subsec:R} to show that there exist SRGs that a particular $p$-particle walk cannot distinguish.
\subsection{Bounding the number of widgets in the non-interacting $p$-particle walk}
Here, we show that $\log_2 X_p \sim p^2$, where $X_p$ is the number of distinct widgets for the non-interacting $p$-boson walk.  First, Auluck proved there are $e^{O(p^{2/3})}$ widgets with no edges \cite{Auluck1953}.  (He counted bipartitions of $(p,q)$, which may be considered to be edgeless widgets when $p=q$.) Since there are at most $2^{p^2}$ ways to add edges to one of these, we have the upper bound $X_p \leq 2^{p^2 + O(p^{2/3})}$.  To get a lower bound, it will suffice to consider the widgets with $2p$ distinct indices.  The edges in one of these can be specified by a $p \times p$ array of bits, and the widgets isomorphic to it are obtained by permuting rows, permuting columns, or transposing the matrix.  Therefore, by Burnside's counting lemma \cite{Liu1968}, the number of isomorphism classes of widgets of this type is
$$
{1 \over |F|} \sum_{f \in F} [\hbox{ \# of arrays fixed by $f$ }],
$$
where the finite group $F$ is the semidirect product of $S_p \times S_p$ by $S_2$.  ($S_p$ and $S_2$ are the symmetric groups on $p$ and 2 objects, respectively.)  This is lower bounded by the term coming from $f=1$, which is $2^{p^2}/( 2 (p!)^2))$, and this is $2^{p^2 + O(p \log p)}$ by Stirling's formula.  From these two estimates the result of Equation~\eqref{eq:log2Xp} follows.
\begin{figure}[h!]
\begin{center}
\includegraphics[width = 7 cm]{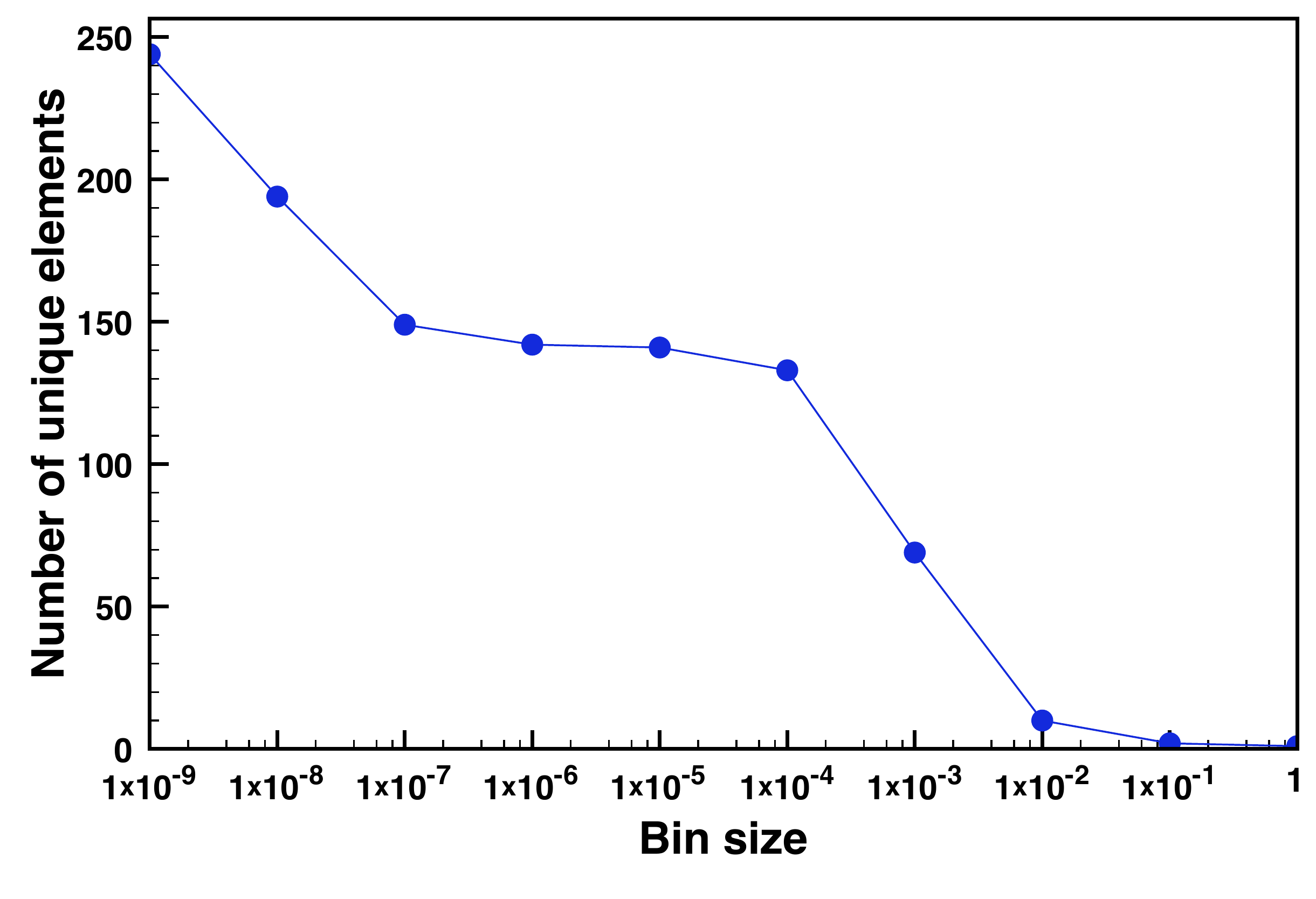}
\caption{\label{fig:S-curve}The number of numerically distinguished elements in the evolution operator ${\bf U}(t)$, defined in Equation~\eqref{eq:U_first} as a function of the bin size used in grouping the elements.  This plot is for the non-interacting three-fermion walk on a graph in the SRG family (16,6,2,2).  We see that the actual number of unique elements is about 150, which can be obtained by using a bin size in the range of $10^{-7}$ to $10^{-4}$.}
\end{center}
\end{figure}
\addtocounter{subfigure}{1}
\subsection{Error analysis for numerical computations}
\label{app:4}
When comparing two graphs, we compute $\Delta$, a measure of the distance between the lists of matrix elements of the evolution operators for the two graphs, as defined in Eq.~\eqref{eq:Delta}.  Computing $\Delta$ requires comparing two lists of numbers that are each exponentially large in particle number $p$.  An evolution operator for a walk of $p$ non-interacting fermions on a graph with $N$ vertices has ${N \choose p}^2$ elements, and the boson equivalent has ${{N+p-1} \choose p}^2$ elements.  For example, the evolution operator for the non-interacting four-fermion walk on a graph of 35 vertices has over 2.7 billion elements.

The comparison of the lists can be made much more efficient by exploiting the fact that the values in the list are highly degenerate.  Instead of comparing the entries in a list, we make histograms of element values and their multiplicities.  We then compute $\Delta$ by comparing these histograms.  When constructing the histograms, it is important to determine the correct bin size.  Choosing too large a bin size results in falsely grouping distinct elements together, while choosing too small a bin size results in falsely distinguishing elements.  By constructing a series of histograms with different bin sizes for the same evolution operator, we are able to determine a range of bin sizes which are neither too large nor too small.  This is illustrated in Figure \ref{fig:S-curve}, which shows that for the non-interacting three-fermion walk on a graph in $(16,6,2,2)$, an appropriate bin size is between $10^{-7}$ and $10^{-4}$.

Because we compute $\Delta$ via numerical simulation, we expect there to be some numerical noise floor.  That is, for any two permutations of the same graph, we expect\\ $\Delta>0$.  It is important to determine how big this quantity, which we denote $\Delta_{iso}$, will be.  We only consider two non-isomorphic graphs to be distinguished if they yield a $\Delta$ which satisfies $\Delta \gg \Delta_{iso}$.  

We numerically compute $\Delta_{iso}$ using double precision arithmetic for a variety of random permutations on our graphs, and we find a maximum $\Delta_{iso}$ to be approximately $10^{-6}$.  Thus, only graph pairs which yield a $\Delta > 10^{-6}$ are considered distinguished.  We find $\Delta_{iso}$ to be relatively insensitive to graph size and particle number.  

In practice, we see a gap for $\Delta$ between distinguished graph pairs and non-distinguished graph pairs.  For distinguished graphs, we find $\Delta$ at least two orders of magnitude larger than $\Delta_{iso}$ (usually much larger); non-distinguished graph pairs have values of $\Delta$ are approximately equal to $\Delta_{iso}$ or are even smaller than it.  
\bibliography{multiparticle}

\begin{thebibliography}{46}%
\makeatletter
\providecommand \@ifxundefined [1]{%
 \@ifx{#1\undefined}
}%
\providecommand \@ifnum [1]{%
 \ifnum #1\expandafter \@firstoftwo
 \else \expandafter \@secondoftwo
 \fi
}%
\providecommand \@ifx [1]{%
 \ifx #1\expandafter \@firstoftwo
 \else \expandafter \@secondoftwo
 \fi
}%
\providecommand \natexlab [1]{#1}%
\providecommand \enquote  [1]{``#1''}%
\providecommand \bibnamefont  [1]{#1}%
\providecommand \bibfnamefont [1]{#1}%
\providecommand \citenamefont [1]{#1}%
\providecommand \href@noop [0]{\@secondoftwo}%
\providecommand \href [0]{\begingroup \@sanitize@url \@href}%
\providecommand \@href[1]{\@@startlink{#1}\@@href}%
\providecommand \@@href[1]{\endgroup#1\@@endlink}%
\providecommand \@sanitize@url [0]{\catcode `\\12\catcode `\$12\catcode
  `\&12\catcode `\#12\catcode `\^12\catcode `\_12\catcode `\%12\relax}%
\providecommand \@@startlink[1]{}%
\providecommand \@@endlink[0]{}%
\providecommand \url  [0]{\begingroup\@sanitize@url \@url }%
\providecommand \@url [1]{\endgroup\@href {#1}{\urlprefix }}%
\providecommand \urlprefix  [0]{URL }%
\providecommand \Eprint [0]{\href }%
\providecommand \doibase [0]{http://dx.doi.org/}%
\providecommand \selectlanguage [0]{\@gobble}%
\providecommand \bibinfo  [0]{\@secondoftwo}%
\providecommand \bibfield  [0]{\@secondoftwo}%
\providecommand \translation [1]{[#1]}%
\providecommand \BibitemOpen [0]{}%
\providecommand \bibitemStop [0]{}%
\providecommand \bibitemNoStop [0]{.\EOS\space}%
\providecommand \EOS [0]{\spacefactor3000\relax}%
\providecommand \BibitemShut  [1]{\csname bibitem#1\endcsname}%
\let\auto@bib@innerbib\@empty
\bibitem [{\citenamefont {Motwani}\ and\ \citenamefont
  {Raghavan}(1996)}]{Motwani1996}%
  \BibitemOpen
  \bibfield  {author} {\bibinfo {author} {\bibfnamefont {R.}~\bibnamefont
  {Motwani}}\ and\ \bibinfo {author} {\bibfnamefont {P.}~\bibnamefont
  {Raghavan}},\ }\href {\doibase http://doi.acm.org/10.1145/234313.234327}
  {\bibfield  {journal} {\bibinfo  {journal} {ACM Comput. Surv.}\ }\textbf
  {\bibinfo {volume} {28}},\ \bibinfo {pages} {33} (\bibinfo {year}
  {1996})}\BibitemShut {NoStop}%
\bibitem [{\citenamefont {Aleliunas}\ \emph {et~al.}(1979)\citenamefont
  {Aleliunas}, \citenamefont {Karp}, \citenamefont {Lipton}, \citenamefont
  {Lovasz},\ and\ \citenamefont {Rackoff}}]{Aleliunas1979}%
  \BibitemOpen
  \bibfield  {author} {\bibinfo {author} {\bibfnamefont {R.}~\bibnamefont
  {Aleliunas}}, \bibinfo {author} {\bibfnamefont {R.~M.}\ \bibnamefont {Karp}},
  \bibinfo {author} {\bibfnamefont {R.~J.}\ \bibnamefont {Lipton}}, \bibinfo
  {author} {\bibfnamefont {L.}~\bibnamefont {Lovasz}}, \ and\ \bibinfo {author}
  {\bibfnamefont {C.}~\bibnamefont {Rackoff}},\ }in\ \href {\doibase
  http://dx.doi.org/10.1109/SFCS.1979.34} {\emph {\bibinfo {booktitle} {FOCS
  '79: Proceedings of the 20th Annual Symposium on Foundations of Computer
  Science}}}\ (\bibinfo  {publisher} {IEEE Computer Society},\ \bibinfo
  {address} {Washington, DC, USA},\ \bibinfo {year} {1979})\ pp.\ \bibinfo
  {pages} {218--223}\BibitemShut {NoStop}%
\bibitem [{\citenamefont {Trautt}\ \emph {et~al.}(2006)\citenamefont {Trautt},
  \citenamefont {Upmanyu},\ and\ \citenamefont {Karma}}]{Trautt2006}%
  \BibitemOpen
  \bibfield  {author} {\bibinfo {author} {\bibfnamefont {Z.}~\bibnamefont
  {Trautt}}, \bibinfo {author} {\bibfnamefont {M.}~\bibnamefont {Upmanyu}}, \
  and\ \bibinfo {author} {\bibfnamefont {A.}~\bibnamefont {Karma}},\ }\href
  {\doibase 10.1126/science.1131988} {\bibfield  {journal} {\bibinfo  {journal}
  {Science}\ }\textbf {\bibinfo {volume} {314}},\ \bibinfo {pages} {632}
  (\bibinfo {year} {2006})}\BibitemShut {NoStop}%
\bibitem [{\citenamefont {Sessions}\ \emph {et~al.}(1997)\citenamefont
  {Sessions}, \citenamefont {Oram},\ and\ \citenamefont
  {Szczelkun}}]{Sessions1997}%
  \BibitemOpen
  \bibfield  {author} {\bibinfo {author} {\bibfnamefont {R.}~\bibnamefont
  {Sessions}}, \bibinfo {author} {\bibfnamefont {M.}~\bibnamefont {Oram}}, \
  and\ \bibinfo {author} {\bibfnamefont {M.}~\bibnamefont {Szczelkun}},\
  }\href@noop {} {\bibfield  {journal} {\bibinfo  {journal} {J. Mol. Biol.}\
  }\textbf {\bibinfo {volume} {270}},\ \bibinfo {pages} {413} (\bibinfo {year}
  {1997})}\BibitemShut {NoStop}%
\bibitem [{\citenamefont {Kilian}\ and\ \citenamefont
  {Taylor}(2003)}]{Kilian2003}%
  \BibitemOpen
  \bibfield  {author} {\bibinfo {author} {\bibfnamefont {L.}~\bibnamefont
  {Kilian}}\ and\ \bibinfo {author} {\bibfnamefont {M.~P.}\ \bibnamefont
  {Taylor}},\ }\href {\doibase DOI: 10.1016/S0022-1996(02)00060-0} {\bibfield
  {journal} {\bibinfo  {journal} {J. Int. Econ.}\ }\textbf {\bibinfo {volume}
  {60}},\ \bibinfo {pages} {85 } (\bibinfo {year} {2003})}\BibitemShut
  {NoStop}%
\bibitem [{\citenamefont {Aharonov}\ \emph {et~al.}(1993)\citenamefont
  {Aharonov}, \citenamefont {Davidovich},\ and\ \citenamefont
  {Zagury}}]{Aharonov1993}%
  \BibitemOpen
  \bibfield  {author} {\bibinfo {author} {\bibfnamefont {Y.}~\bibnamefont
  {Aharonov}}, \bibinfo {author} {\bibfnamefont {L.}~\bibnamefont
  {Davidovich}}, \ and\ \bibinfo {author} {\bibfnamefont {N.}~\bibnamefont
  {Zagury}},\ }\href {\doibase 10.1103/PhysRevA.48.1687} {\bibfield  {journal}
  {\bibinfo  {journal} {Phys. Rev. A}\ }\textbf {\bibinfo {volume} {48}},\
  \bibinfo {pages} {1687} (\bibinfo {year} {1993})}\BibitemShut {NoStop}%
\bibitem [{\citenamefont {Bach}\ \emph {et~al.}(2004)\citenamefont {Bach},
  \citenamefont {Coppersmith}, \citenamefont {Goldschen}, \citenamefont
  {Joynt},\ and\ \citenamefont {Watrous}}]{Bach2004}%
  \BibitemOpen
  \bibfield  {author} {\bibinfo {author} {\bibfnamefont {E.}~\bibnamefont
  {Bach}}, \bibinfo {author} {\bibfnamefont {S.}~\bibnamefont {Coppersmith}},
  \bibinfo {author} {\bibfnamefont {M.~P.}\ \bibnamefont {Goldschen}}, \bibinfo
  {author} {\bibfnamefont {R.}~\bibnamefont {Joynt}}, \ and\ \bibinfo {author}
  {\bibfnamefont {J.}~\bibnamefont {Watrous}},\ }\href {\doibase DOI:
  10.1016/j.jcss.2004.03.005} {\bibfield  {journal} {\bibinfo  {journal}
  {Journal of Computer and System Sciences}\ }\textbf {\bibinfo {volume}
  {69}},\ \bibinfo {pages} {562 } (\bibinfo {year} {2004})}\BibitemShut
  {NoStop}%
\bibitem [{\citenamefont {Solenov}\ and\ \citenamefont
  {Fedichkin}(2006)}]{Solenov2006}%
  \BibitemOpen
  \bibfield  {author} {\bibinfo {author} {\bibfnamefont {D.}~\bibnamefont
  {Solenov}}\ and\ \bibinfo {author} {\bibfnamefont {L.}~\bibnamefont
  {Fedichkin}},\ }\href {\doibase 10.1103/PhysRevA.73.012313} {\bibfield
  {journal} {\bibinfo  {journal} {Phys. Rev. A}\ }\textbf {\bibinfo {volume}
  {73}},\ \bibinfo {pages} {012313} (\bibinfo {year} {2006})}\BibitemShut
  {NoStop}%
\bibitem [{\citenamefont {Childs}\ \emph {et~al.}(2002)\citenamefont {Childs},
  \citenamefont {Farhi},\ and\ \citenamefont {Gutmann}}]{Childs2002}%
  \BibitemOpen
  \bibfield  {author} {\bibinfo {author} {\bibfnamefont {A.}~\bibnamefont
  {Childs}}, \bibinfo {author} {\bibfnamefont {E.}~\bibnamefont {Farhi}}, \
  and\ \bibinfo {author} {\bibfnamefont {S.}~\bibnamefont {Gutmann}},\
  }\href@noop {} {\bibfield  {journal} {\bibinfo  {journal} {Quantum Inf.
  Process.}\ } (\bibinfo {year} {2002})}\BibitemShut {NoStop}%
\bibitem [{\citenamefont {Shenvi}\ \emph {et~al.}(2003)\citenamefont {Shenvi},
  \citenamefont {Kempe},\ and\ \citenamefont {Whaley}}]{Shenvi2003}%
  \BibitemOpen
  \bibfield  {author} {\bibinfo {author} {\bibfnamefont {N.}~\bibnamefont
  {Shenvi}}, \bibinfo {author} {\bibfnamefont {J.}~\bibnamefont {Kempe}}, \
  and\ \bibinfo {author} {\bibfnamefont {K.~B.}\ \bibnamefont {Whaley}},\
  }\href {\doibase 10.1103/PhysRevA.67.052307} {\bibfield  {journal} {\bibinfo
  {journal} {Phys. Rev. A}\ }\textbf {\bibinfo {volume} {67}},\ \bibinfo
  {pages} {052307} (\bibinfo {year} {2003})}\BibitemShut {NoStop}%
\bibitem [{\citenamefont {Ambainis}(2003)}]{Ambainis2003}%
  \BibitemOpen
  \bibfield  {author} {\bibinfo {author} {\bibfnamefont {A.}~\bibnamefont
  {Ambainis}},\ }\href@noop {} {\bibfield  {journal} {\bibinfo  {journal}
  {International Journal of Quantum Information}\ }\textbf {\bibinfo {volume}
  {1}},\ \bibinfo {pages} {507} (\bibinfo {year} {2003})}\BibitemShut {NoStop}%
\bibitem [{\citenamefont {Ambainis}(2004)}]{Ambainis2004}%
  \BibitemOpen
  \bibfield  {author} {\bibinfo {author} {\bibfnamefont {A.}~\bibnamefont
  {Ambainis}},\ }\href {\doibase
  http://doi.ieeecomputersociety.org/10.1109/FOCS.2004.54} {\bibfield
  {journal} {\bibinfo  {journal} {Foundations of Computer Science, Annual IEEE
  Symposium on}\ }\textbf {\bibinfo {volume} {0}},\ \bibinfo {pages} {22}
  (\bibinfo {year} {2004})}\BibitemShut {NoStop}%
\bibitem [{\citenamefont {Magniez}\ \emph {et~al.}(2007)\citenamefont
  {Magniez}, \citenamefont {Nayak}, \citenamefont {Roland},\ and\ \citenamefont
  {Santha}}]{Magniez2007}%
  \BibitemOpen
  \bibfield  {author} {\bibinfo {author} {\bibfnamefont {F.}~\bibnamefont
  {Magniez}}, \bibinfo {author} {\bibfnamefont {A.}~\bibnamefont {Nayak}},
  \bibinfo {author} {\bibfnamefont {J.}~\bibnamefont {Roland}}, \ and\ \bibinfo
  {author} {\bibfnamefont {M.}~\bibnamefont {Santha}},\ }in\ \href {\doibase
  http://doi.acm.org/10.1145/1250790.1250874} {\emph {\bibinfo {booktitle}
  {STOC '07: Proceedings of the thirty-ninth annual ACM symposium on Theory of
  computing}}}\ (\bibinfo  {publisher} {ACM},\ \bibinfo {address} {New York,
  NY, USA},\ \bibinfo {year} {2007})\ pp.\ \bibinfo {pages}
  {575--584}\BibitemShut {NoStop}%
\bibitem [{\citenamefont {Poto\ifmmode~\check{c}\else \v{c}\fi{}ek}\ \emph
  {et~al.}(2009)\citenamefont {Poto\ifmmode~\check{c}\else \v{c}\fi{}ek},
  \citenamefont {G\'abris}, \citenamefont {Kiss},\ and\ \citenamefont
  {Jex}}]{Potocek2009}%
  \BibitemOpen
  \bibfield  {author} {\bibinfo {author} {\bibfnamefont {V.}~\bibnamefont
  {Poto\ifmmode~\check{c}\else \v{c}\fi{}ek}}, \bibinfo {author} {\bibfnamefont
  {A.}~\bibnamefont {G\'abris}}, \bibinfo {author} {\bibfnamefont
  {T.}~\bibnamefont {Kiss}}, \ and\ \bibinfo {author} {\bibfnamefont
  {I.}~\bibnamefont {Jex}},\ }\href {\doibase 10.1103/PhysRevA.79.012325}
  {\bibfield  {journal} {\bibinfo  {journal} {Phys. Rev. A}\ }\textbf {\bibinfo
  {volume} {79}},\ \bibinfo {pages} {012325} (\bibinfo {year}
  {2009})}\BibitemShut {NoStop}%
\bibitem [{\citenamefont {Reitzner}\ \emph {et~al.}(2009)\citenamefont
  {Reitzner}, \citenamefont {Hillery}, \citenamefont {Feldman},\ and\
  \citenamefont {Bu\ifmmode~\check{z}\else \v{z}\fi{}ek}}]{Reitzner2009}%
  \BibitemOpen
  \bibfield  {author} {\bibinfo {author} {\bibfnamefont {D.}~\bibnamefont
  {Reitzner}}, \bibinfo {author} {\bibfnamefont {M.}~\bibnamefont {Hillery}},
  \bibinfo {author} {\bibfnamefont {E.}~\bibnamefont {Feldman}}, \ and\
  \bibinfo {author} {\bibfnamefont {V.}~\bibnamefont {Bu\ifmmode~\check{z}\else
  \v{z}\fi{}ek}},\ }\href {\doibase 10.1103/PhysRevA.79.012323} {\bibfield
  {journal} {\bibinfo  {journal} {Phys. Rev. A}\ }\textbf {\bibinfo {volume}
  {79}},\ \bibinfo {pages} {012323} (\bibinfo {year} {2009})}\BibitemShut
  {NoStop}%
\bibitem [{\citenamefont {Schmitz}\ \emph {et~al.}(2009)\citenamefont
  {Schmitz}, \citenamefont {Matjeschk}, \citenamefont {Schneider},
  \citenamefont {Glueckert}, \citenamefont {Enderlein}, \citenamefont {Huber},\
  and\ \citenamefont {Schaetz}}]{Schmitz2009}%
  \BibitemOpen
  \bibfield  {author} {\bibinfo {author} {\bibfnamefont {H.}~\bibnamefont
  {Schmitz}}, \bibinfo {author} {\bibfnamefont {R.}~\bibnamefont {Matjeschk}},
  \bibinfo {author} {\bibfnamefont {C.}~\bibnamefont {Schneider}}, \bibinfo
  {author} {\bibfnamefont {J.}~\bibnamefont {Glueckert}}, \bibinfo {author}
  {\bibfnamefont {M.}~\bibnamefont {Enderlein}}, \bibinfo {author}
  {\bibfnamefont {T.}~\bibnamefont {Huber}}, \ and\ \bibinfo {author}
  {\bibfnamefont {T.}~\bibnamefont {Schaetz}},\ }\href {\doibase
  10.1103/PhysRevLett.103.090504} {\bibfield  {journal} {\bibinfo  {journal}
  {Phys. Rev. Lett.}\ }\textbf {\bibinfo {volume} {103}},\ \bibinfo {pages}
  {090504} (\bibinfo {year} {2009})}\BibitemShut {NoStop}%
\bibitem [{\citenamefont {Karski}\ \emph {et~al.}(2009)\citenamefont {Karski},
  \citenamefont {Fšrster}, \citenamefont {Choi}, \citenamefont {Steffen},
  \citenamefont {Alt}, \citenamefont {Meschede},\ and\ \citenamefont
  {Widera}}]{Karski2009}%
  \BibitemOpen
  \bibfield  {author} {\bibinfo {author} {\bibfnamefont {M.}~\bibnamefont
  {Karski}}, \bibinfo {author} {\bibfnamefont {L.}~\bibnamefont {Fšrster}},
  \bibinfo {author} {\bibfnamefont {J.-M.}\ \bibnamefont {Choi}}, \bibinfo
  {author} {\bibfnamefont {A.}~\bibnamefont {Steffen}}, \bibinfo {author}
  {\bibfnamefont {W.}~\bibnamefont {Alt}}, \bibinfo {author} {\bibfnamefont
  {D.}~\bibnamefont {Meschede}}, \ and\ \bibinfo {author} {\bibfnamefont
  {A.}~\bibnamefont {Widera}},\ }\href {\doibase 10.1126/science.1174436}
  {\bibfield  {journal} {\bibinfo  {journal} {Science}\ }\textbf {\bibinfo
  {volume} {325}},\ \bibinfo {pages} {174} (\bibinfo {year}
  {2009})}\BibitemShut {NoStop}%
\bibitem [{\citenamefont {Schreiber}\ \emph {et~al.}(2010)\citenamefont
  {Schreiber}, \citenamefont {Cassemiro}, \citenamefont
  {Poto\ifmmode~\check{c}\else \v{c}\fi{}ek}, \citenamefont {G\'abris},
  \citenamefont {Mosley}, \citenamefont {Andersson}, \citenamefont {Jex},\ and\
  \citenamefont {Silberhorn}}]{Schreiber2010}%
  \BibitemOpen
  \bibfield  {author} {\bibinfo {author} {\bibfnamefont {A.}~\bibnamefont
  {Schreiber}}, \bibinfo {author} {\bibfnamefont {K.~N.}\ \bibnamefont
  {Cassemiro}}, \bibinfo {author} {\bibfnamefont {V.}~\bibnamefont
  {Poto\ifmmode~\check{c}\else \v{c}\fi{}ek}}, \bibinfo {author} {\bibfnamefont
  {A.}~\bibnamefont {G\'abris}}, \bibinfo {author} {\bibfnamefont {P.~J.}\
  \bibnamefont {Mosley}}, \bibinfo {author} {\bibfnamefont {E.}~\bibnamefont
  {Andersson}}, \bibinfo {author} {\bibfnamefont {I.}~\bibnamefont {Jex}}, \
  and\ \bibinfo {author} {\bibfnamefont {C.}~\bibnamefont {Silberhorn}},\
  }\href {\doibase 10.1103/PhysRevLett.104.050502} {\bibfield  {journal}
  {\bibinfo  {journal} {Phys. Rev. Lett.}\ }\textbf {\bibinfo {volume} {104}},\
  \bibinfo {pages} {050502} (\bibinfo {year} {2010})}\BibitemShut {NoStop}%
\bibitem [{\citenamefont {Broome}\ \emph {et~al.}(2010)\citenamefont {Broome},
  \citenamefont {Fedrizzi}, \citenamefont {Lanyon}, \citenamefont {Kassal},
  \citenamefont {Aspuru-Guzik},\ and\ \citenamefont {White}}]{Broome2010}%
  \BibitemOpen
  \bibfield  {author} {\bibinfo {author} {\bibfnamefont {M.~A.}\ \bibnamefont
  {Broome}}, \bibinfo {author} {\bibfnamefont {A.}~\bibnamefont {Fedrizzi}},
  \bibinfo {author} {\bibfnamefont {B.~P.}\ \bibnamefont {Lanyon}}, \bibinfo
  {author} {\bibfnamefont {I.}~\bibnamefont {Kassal}}, \bibinfo {author}
  {\bibfnamefont {A.}~\bibnamefont {Aspuru-Guzik}}, \ and\ \bibinfo {author}
  {\bibfnamefont {A.~G.}\ \bibnamefont {White}},\ }\href {\doibase
  10.1103/PhysRevLett.104.153602} {\bibfield  {journal} {\bibinfo  {journal}
  {Phys. Rev. Lett.}\ }\textbf {\bibinfo {volume} {104}},\ \bibinfo {pages}
  {153602} (\bibinfo {year} {2010})}\BibitemShut {NoStop}%
\bibitem [{\citenamefont {Ryan}\ \emph {et~al.}(2005)\citenamefont {Ryan},
  \citenamefont {Laforest}, \citenamefont {Boileau},\ and\ \citenamefont
  {Laflamme}}]{Ryan2005}%
  \BibitemOpen
  \bibfield  {author} {\bibinfo {author} {\bibfnamefont {C.~A.}\ \bibnamefont
  {Ryan}}, \bibinfo {author} {\bibfnamefont {M.}~\bibnamefont {Laforest}},
  \bibinfo {author} {\bibfnamefont {J.~C.}\ \bibnamefont {Boileau}}, \ and\
  \bibinfo {author} {\bibfnamefont {R.}~\bibnamefont {Laflamme}},\ }\href
  {\doibase 10.1103/PhysRevA.72.062317} {\bibfield  {journal} {\bibinfo
  {journal} {Phys. Rev. A}\ }\textbf {\bibinfo {volume} {72}},\ \bibinfo
  {pages} {062317} (\bibinfo {year} {2005})}\BibitemShut {NoStop}%
\bibitem [{\citenamefont {Z\"ahringer}\ \emph {et~al.}(2010)\citenamefont
  {Z\"ahringer}, \citenamefont {Kirchmair}, \citenamefont {Gerritsma},
  \citenamefont {Solano}, \citenamefont {Blatt},\ and\ \citenamefont
  {Roos}}]{Zahringer2010}%
  \BibitemOpen
  \bibfield  {author} {\bibinfo {author} {\bibfnamefont {F.}~\bibnamefont
  {Z\"ahringer}}, \bibinfo {author} {\bibfnamefont {G.}~\bibnamefont
  {Kirchmair}}, \bibinfo {author} {\bibfnamefont {R.}~\bibnamefont
  {Gerritsma}}, \bibinfo {author} {\bibfnamefont {E.}~\bibnamefont {Solano}},
  \bibinfo {author} {\bibfnamefont {R.}~\bibnamefont {Blatt}}, \ and\ \bibinfo
  {author} {\bibfnamefont {C.~F.}\ \bibnamefont {Roos}},\ }\href {\doibase
  10.1103/PhysRevLett.104.100503} {\bibfield  {journal} {\bibinfo  {journal}
  {Phys. Rev. Lett.}\ }\textbf {\bibinfo {volume} {104}},\ \bibinfo {pages}
  {100503} (\bibinfo {year} {2010})}\BibitemShut {NoStop}%
\bibitem [{\citenamefont {Owens}\ \emph {et~al.}(2011)\citenamefont {Owens},
  \citenamefont {Broome}, \citenamefont {Biggerstaff}, \citenamefont {Goggin},
  \citenamefont {Fedrizzi}, \citenamefont {Linjordet}, \citenamefont {Ams},
  \citenamefont {Marshall}, \citenamefont {Twamley}, \citenamefont {Withford},\
  and\ \citenamefont {White}}]{Owens2011}%
  \BibitemOpen
  \bibfield  {author} {\bibinfo {author} {\bibfnamefont {J.~O.}\ \bibnamefont
  {Owens}}, \bibinfo {author} {\bibfnamefont {M.~A.}\ \bibnamefont {Broome}},
  \bibinfo {author} {\bibfnamefont {D.~N.}\ \bibnamefont {Biggerstaff}},
  \bibinfo {author} {\bibfnamefont {M.~E.}\ \bibnamefont {Goggin}}, \bibinfo
  {author} {\bibfnamefont {A.}~\bibnamefont {Fedrizzi}}, \bibinfo {author}
  {\bibfnamefont {T.}~\bibnamefont {Linjordet}}, \bibinfo {author}
  {\bibfnamefont {M.}~\bibnamefont {Ams}}, \bibinfo {author} {\bibfnamefont
  {G.~D.}\ \bibnamefont {Marshall}}, \bibinfo {author} {\bibfnamefont
  {J.}~\bibnamefont {Twamley}}, \bibinfo {author} {\bibfnamefont {M.~J.}\
  \bibnamefont {Withford}}, \ and\ \bibinfo {author} {\bibfnamefont {A.~G.}\
  \bibnamefont {White}},\ }\href
  {http://stacks.iop.org/1367-2630/13/i=7/a=075003} {\bibfield  {journal}
  {\bibinfo  {journal} {New Journal of Physics}\ }\textbf {\bibinfo {volume}
  {13}},\ \bibinfo {pages} {075003} (\bibinfo {year} {2011})}\BibitemShut
  {NoStop}%
\bibitem [{\citenamefont {Peruzzo}\ \emph {et~al.}(2010)\citenamefont
  {Peruzzo}, \citenamefont {Lobino}, \citenamefont {Matthews}, \citenamefont
  {Matsuda}, \citenamefont {Politi}, \citenamefont {Poulios}, \citenamefont
  {Zhou}, \citenamefont {Lahini}, \citenamefont {Ismail}, \citenamefont
  {Wörhoff}, \citenamefont {Bromberg}, \citenamefont {Silberberg},
  \citenamefont {Thompson},\ and\ \citenamefont {OBrien}}]{Peruzzo2010}%
  \BibitemOpen
  \bibfield  {author} {\bibinfo {author} {\bibfnamefont {A.}~\bibnamefont
  {Peruzzo}}, \bibinfo {author} {\bibfnamefont {M.}~\bibnamefont {Lobino}},
  \bibinfo {author} {\bibfnamefont {J.~C.~F.}\ \bibnamefont {Matthews}},
  \bibinfo {author} {\bibfnamefont {N.}~\bibnamefont {Matsuda}}, \bibinfo
  {author} {\bibfnamefont {A.}~\bibnamefont {Politi}}, \bibinfo {author}
  {\bibfnamefont {K.}~\bibnamefont {Poulios}}, \bibinfo {author} {\bibfnamefont
  {X.-Q.}\ \bibnamefont {Zhou}}, \bibinfo {author} {\bibfnamefont
  {Y.}~\bibnamefont {Lahini}}, \bibinfo {author} {\bibfnamefont
  {N.}~\bibnamefont {Ismail}}, \bibinfo {author} {\bibfnamefont
  {K.}~\bibnamefont {Wörhoff}}, \bibinfo {author} {\bibfnamefont
  {Y.}~\bibnamefont {Bromberg}}, \bibinfo {author} {\bibfnamefont
  {Y.}~\bibnamefont {Silberberg}}, \bibinfo {author} {\bibfnamefont {M.~G.}\
  \bibnamefont {Thompson}}, \ and\ \bibinfo {author} {\bibfnamefont {J.~L.}\
  \bibnamefont {OBrien}},\ }\href {\doibase 10.1126/science.1193515} {\bibfield
   {journal} {\bibinfo  {journal} {Science}\ }\textbf {\bibinfo {volume}
  {329}},\ \bibinfo {pages} {1500} (\bibinfo {year} {2010})}\BibitemShut
  {NoStop}%
\bibitem [{\citenamefont {Sansoni}\ \emph {et~al.}(2012)\citenamefont
  {Sansoni}, \citenamefont {Sciarrino}, \citenamefont {Vallone}, \citenamefont
  {Mataloni}, \citenamefont {Crespi}, \citenamefont {Ramponi},\ and\
  \citenamefont {Osellame}}]{Sansoni2012}%
  \BibitemOpen
  \bibfield  {author} {\bibinfo {author} {\bibfnamefont {L.}~\bibnamefont
  {Sansoni}}, \bibinfo {author} {\bibfnamefont {F.}~\bibnamefont {Sciarrino}},
  \bibinfo {author} {\bibfnamefont {G.}~\bibnamefont {Vallone}}, \bibinfo
  {author} {\bibfnamefont {P.}~\bibnamefont {Mataloni}}, \bibinfo {author}
  {\bibfnamefont {A.}~\bibnamefont {Crespi}}, \bibinfo {author} {\bibfnamefont
  {R.}~\bibnamefont {Ramponi}}, \ and\ \bibinfo {author} {\bibfnamefont
  {R.}~\bibnamefont {Osellame}},\ }\href {\doibase
  10.1103/PhysRevLett.108.010502} {\bibfield  {journal} {\bibinfo  {journal}
  {Phys. Rev. Lett.}\ }\textbf {\bibinfo {volume} {108}},\ \bibinfo {pages}
  {010502} (\bibinfo {year} {2012})}\BibitemShut {NoStop}%
\bibitem [{\citenamefont {Manouchehri}\ and\ \citenamefont
  {Wang}(2009)}]{Manouchehri2009}%
  \BibitemOpen
  \bibfield  {author} {\bibinfo {author} {\bibfnamefont {K.}~\bibnamefont
  {Manouchehri}}\ and\ \bibinfo {author} {\bibfnamefont {J.~B.}\ \bibnamefont
  {Wang}},\ }\href {\doibase 10.1103/PhysRevA.80.060304} {\bibfield  {journal}
  {\bibinfo  {journal} {Phys. Rev. A}\ }\textbf {\bibinfo {volume} {80}},\
  \bibinfo {pages} {060304} (\bibinfo {year} {2009})}\BibitemShut {NoStop}%
\bibitem [{\citenamefont {Childs}(2009)}]{Childs2009}%
  \BibitemOpen
  \bibfield  {author} {\bibinfo {author} {\bibfnamefont {A.~M.}\ \bibnamefont
  {Childs}},\ }\href {\doibase 10.1103/PhysRevLett.102.180501} {\bibfield
  {journal} {\bibinfo  {journal} {Phys. Rev. Lett.}\ }\textbf {\bibinfo
  {volume} {102}},\ \bibinfo {pages} {180501} (\bibinfo {year}
  {2009})}\BibitemShut {NoStop}%
\bibitem [{\citenamefont {Gamble}\ \emph {et~al.}(2010)\citenamefont {Gamble},
  \citenamefont {Friesen}, \citenamefont {Zhou}, \citenamefont {Joynt},\ and\
  \citenamefont {Coppersmith}}]{Gamble2010}%
  \BibitemOpen
  \bibfield  {author} {\bibinfo {author} {\bibfnamefont {J.~K.}\ \bibnamefont
  {Gamble}}, \bibinfo {author} {\bibfnamefont {M.}~\bibnamefont {Friesen}},
  \bibinfo {author} {\bibfnamefont {D.}~\bibnamefont {Zhou}}, \bibinfo {author}
  {\bibfnamefont {R.}~\bibnamefont {Joynt}}, \ and\ \bibinfo {author}
  {\bibfnamefont {S.~N.}\ \bibnamefont {Coppersmith}},\ }\href {\doibase
  10.1103/PhysRevA.81.052313} {\bibfield  {journal} {\bibinfo  {journal} {Phys.
  Rev. A}\ }\textbf {\bibinfo {volume} {81}},\ \bibinfo {pages} {052313}
  (\bibinfo {year} {2010})}\BibitemShut {NoStop}%
\bibitem [{\citenamefont {Burioni}\ \emph {et~al.}(2000)\citenamefont
  {Burioni}, \citenamefont {Cassi}, \citenamefont {Meccoli}, \citenamefont
  {Rasetti}, \citenamefont {Regina}, \citenamefont {Sodano},\ and\
  \citenamefont {Vezzani}}]{Burioni2000}%
  \BibitemOpen
  \bibfield  {author} {\bibinfo {author} {\bibfnamefont {R.}~\bibnamefont
  {Burioni}}, \bibinfo {author} {\bibfnamefont {D.}~\bibnamefont {Cassi}},
  \bibinfo {author} {\bibfnamefont {I.}~\bibnamefont {Meccoli}}, \bibinfo
  {author} {\bibfnamefont {M.}~\bibnamefont {Rasetti}}, \bibinfo {author}
  {\bibfnamefont {S.}~\bibnamefont {Regina}}, \bibinfo {author} {\bibfnamefont
  {P.}~\bibnamefont {Sodano}}, \ and\ \bibinfo {author} {\bibfnamefont
  {A.}~\bibnamefont {Vezzani}},\ }\href {\doibase 10.1209/epl/i2000-00431-5}
  {\bibfield  {journal} {\bibinfo  {journal} {Europhys. Lett.}\ }\textbf
  {\bibinfo {volume} {52}},\ \bibinfo {pages} {251} (\bibinfo {year}
  {2000})}\BibitemShut {NoStop}%
\bibitem [{\citenamefont {Buonsante}\ \emph {et~al.}(2002)\citenamefont
  {Buonsante}, \citenamefont {Burioni}, \citenamefont {Cassi},\ and\
  \citenamefont {Vezzani}}]{Buonsante2002}%
  \BibitemOpen
  \bibfield  {author} {\bibinfo {author} {\bibfnamefont {P.}~\bibnamefont
  {Buonsante}}, \bibinfo {author} {\bibfnamefont {R.}~\bibnamefont {Burioni}},
  \bibinfo {author} {\bibfnamefont {D.}~\bibnamefont {Cassi}}, \ and\ \bibinfo
  {author} {\bibfnamefont {A.}~\bibnamefont {Vezzani}},\ }\href {\doibase
  10.1103/PhysRevB.66.094207} {\bibfield  {journal} {\bibinfo  {journal} {Phys.
  Rev. B}\ }\textbf {\bibinfo {volume} {66}},\ \bibinfo {pages} {094207}
  (\bibinfo {year} {2002})}\BibitemShut {NoStop}%
\bibitem [{\citenamefont {Mancini}\ \emph {et~al.}(2007)\citenamefont
  {Mancini}, \citenamefont {Sodano},\ and\ \citenamefont
  {Trombettoni}}]{Mancini2007}%
  \BibitemOpen
  \bibfield  {author} {\bibinfo {author} {\bibfnamefont {F.~P.}\ \bibnamefont
  {Mancini}}, \bibinfo {author} {\bibfnamefont {P.}~\bibnamefont {Sodano}}, \
  and\ \bibinfo {author} {\bibfnamefont {A.}~\bibnamefont {Trombettoni}},\
  }\href@noop {} {\bibfield  {journal} {\bibinfo  {journal} {AIP Conference
  Proceedings}\ }\textbf {\bibinfo {volume} {918}},\ \bibinfo {pages} {302}
  (\bibinfo {year} {2007})}\BibitemShut {NoStop}%
\bibitem [{\citenamefont {Spielman}(1996)}]{Spielman1996}%
  \BibitemOpen
  \bibfield  {author} {\bibinfo {author} {\bibfnamefont {D.~A.}\ \bibnamefont
  {Spielman}},\ }in\ \href {\doibase http://doi.acm.org/10.1145/237814.238006}
  {\emph {\bibinfo {booktitle} {Proceedings of the twenty-eighth annual ACM
  symposium on Theory of computing}}},\ \bibinfo {series and number} {STOC
  '96}\ (\bibinfo  {publisher} {ACM},\ \bibinfo {address} {New York, NY, USA},\
  \bibinfo {year} {1996})\ pp.\ \bibinfo {pages} {576--584}\BibitemShut
  {NoStop}%
\bibitem [{\citenamefont {Sch{\"o}ning}(1988)}]{Schoning1988}%
  \BibitemOpen
  \bibfield  {author} {\bibinfo {author} {\bibfnamefont {U.}~\bibnamefont
  {Sch{\"o}ning}},\ }\href@noop {} {\bibfield  {journal} {\bibinfo  {journal}
  {J. Comp. Syst. Sci.}\ }\textbf {\bibinfo {volume} {37}},\ \bibinfo {pages}
  {312} (\bibinfo {year} {1988})}\BibitemShut {NoStop}%
\bibitem [{\citenamefont {Moore}\ \emph {et~al.}(2007)\citenamefont {Moore},
  \citenamefont {Russell},\ and\ \citenamefont {Sniady}}]{Moore2007}%
  \BibitemOpen
  \bibfield  {author} {\bibinfo {author} {\bibfnamefont {C.}~\bibnamefont
  {Moore}}, \bibinfo {author} {\bibfnamefont {A.}~\bibnamefont {Russell}}, \
  and\ \bibinfo {author} {\bibfnamefont {P.}~\bibnamefont {Sniady}},\ }in\
  \href {\doibase http://doi.acm.org/10.1145/1250790.1250868} {\emph {\bibinfo
  {booktitle} {STOC '07: Proceedings of the thirty-ninth annual ACM symposium
  on Theory of computing}}}\ (\bibinfo  {publisher} {ACM},\ \bibinfo {address}
  {New York, NY, USA},\ \bibinfo {year} {2007})\ pp.\ \bibinfo {pages}
  {536--545}\BibitemShut {NoStop}%
\bibitem [{\citenamefont {Shor}(1994)}]{Shor1994}%
  \BibitemOpen
  \bibfield  {author} {\bibinfo {author} {\bibfnamefont {P.~W.}\ \bibnamefont
  {Shor}},\ }in\ \href {\doibase http://dx.doi.org/10.1109/SFCS.1994.365700}
  {\emph {\bibinfo {booktitle} {FOCS '94: Proceedings of the 35th Annual
  Symposium on Foundations of Computer Science}}}\ (\bibinfo  {publisher} {IEEE
  Computer Society},\ \bibinfo {address} {Washington, DC, USA},\ \bibinfo
  {year} {1994})\ pp.\ \bibinfo {pages} {124--134}\BibitemShut {NoStop}%
\bibitem [{\citenamefont {Shiau}\ \emph {et~al.}(2005)\citenamefont {Shiau},
  \citenamefont {Joynt},\ and\ \citenamefont {Coppersmith}}]{Shiau2005}%
  \BibitemOpen
  \bibfield  {author} {\bibinfo {author} {\bibfnamefont {S.}~\bibnamefont
  {Shiau}}, \bibinfo {author} {\bibfnamefont {R.}~\bibnamefont {Joynt}}, \ and\
  \bibinfo {author} {\bibfnamefont {S.}~\bibnamefont {Coppersmith}},\
  }\href@noop {} {\bibfield  {journal} {\bibinfo  {journal} {Quantum Inf.
  Comput.}\ }\textbf {\bibinfo {volume} {5}},\ \bibinfo {pages} {492} (\bibinfo
  {year} {2005})}\BibitemShut {NoStop}%
\bibitem [{\citenamefont {Smith}(2010)}]{JSmith2010}%
  \BibitemOpen
  \bibfield  {author} {\bibinfo {author} {\bibfnamefont {J.}~\bibnamefont
  {Smith}},\ }\href@noop {} {\  (\bibinfo {year} {2010})},\ \Eprint
  {http://arxiv.org/abs/1004.0206v1} {arXiv:1004.0206v1 [quant-ph]}
  \BibitemShut {NoStop}%
\bibitem [{\citenamefont {Emms}\ \emph {et~al.}(2009)\citenamefont {Emms},
  \citenamefont {Severini}, \citenamefont {Wilson},\ and\ \citenamefont
  {Hancock}}]{Emms2009}%
  \BibitemOpen
  \bibfield  {author} {\bibinfo {author} {\bibfnamefont {D.}~\bibnamefont
  {Emms}}, \bibinfo {author} {\bibfnamefont {S.}~\bibnamefont {Severini}},
  \bibinfo {author} {\bibfnamefont {R.~C.}\ \bibnamefont {Wilson}}, \ and\
  \bibinfo {author} {\bibfnamefont {E.~R.}\ \bibnamefont {Hancock}},\ }\href
  {\doibase 10.1016/j.patcog.2008.10.025} {\bibfield  {journal} {\bibinfo
  {journal} {Pattern Recogn.}\ }\textbf {\bibinfo {volume} {42}} (\bibinfo
  {year} {2009}),\ 10.1016/j.patcog.2008.10.025}\BibitemShut {NoStop}%
\bibitem [{\citenamefont {Godsil}\ and\ \citenamefont {Guo}(2011)}]{Guo2011}%
  \BibitemOpen
  \bibfield  {author} {\bibinfo {author} {\bibfnamefont {C.}~\bibnamefont
  {Godsil}}\ and\ \bibinfo {author} {\bibfnamefont {K.}~\bibnamefont {Guo}},\
  }\href@noop {} {\bibfield  {journal} {\bibinfo  {journal} {The Electronic
  Journal of Combinatorics}\ }\textbf {\bibinfo {volume} {18}} (\bibinfo {year}
  {2011})}\BibitemShut {NoStop}%
\bibitem [{\citenamefont {Berry}\ and\ \citenamefont {Wang}(2011)}]{Berry2011}%
  \BibitemOpen
  \bibfield  {author} {\bibinfo {author} {\bibfnamefont {S.~D.}\ \bibnamefont
  {Berry}}\ and\ \bibinfo {author} {\bibfnamefont {J.~B.}\ \bibnamefont
  {Wang}},\ }\href {\doibase 10.1103/PhysRevA.83.042317} {\bibfield  {journal}
  {\bibinfo  {journal} {Phys. Rev. A}\ }\textbf {\bibinfo {volume} {83}},\
  \bibinfo {pages} {042317} (\bibinfo {year} {2011})}\BibitemShut {NoStop}%
\bibitem [{\citenamefont {Godsil}\ and\ \citenamefont
  {Royle}(2001)}]{Godsil2001}%
  \BibitemOpen
  \bibfield  {author} {\bibinfo {author} {\bibfnamefont {C.}~\bibnamefont
  {Godsil}}\ and\ \bibinfo {author} {\bibfnamefont {G.}~\bibnamefont {Royle}},\
  }\href@noop {} {\emph {\bibinfo {title} {Algebraic Graph Theory}}}\ (\bibinfo
   {publisher} {Springer},\ \bibinfo {year} {2001})\BibitemShut {NoStop}%
\bibitem [{\citenamefont {Sidje}(1998)}]{expokit}%
  \BibitemOpen
  \bibfield  {author} {\bibinfo {author} {\bibfnamefont {R.~B.}\ \bibnamefont
  {Sidje}},\ }\href@noop {} {\bibfield  {journal} {\bibinfo  {journal} {ACM
  Trans. Math. Softw.}\ }\textbf {\bibinfo {volume} {24}},\ \bibinfo {pages}
  {130} (\bibinfo {year} {1998})}\BibitemShut {NoStop}%
\bibitem [{\citenamefont {Stones}(2010)}]{Stones2010}%
  \BibitemOpen
  \bibfield  {author} {\bibinfo {author} {\bibfnamefont {D.~S.}\ \bibnamefont
  {Stones}},\ }\href@noop {} {\bibfield  {journal} {\bibinfo  {journal}
  {Electr. J. Comb.}\ }\textbf {\bibinfo {volume} {17}} (\bibinfo {year}
  {2010})}\BibitemShut {NoStop}%
\bibitem [{\citenamefont {McKay}\ and\ \citenamefont
  {Wanless}(2005)}]{McKay2005}%
  \BibitemOpen
  \bibfield  {author} {\bibinfo {author} {\bibfnamefont {B.}~\bibnamefont
  {McKay}}\ and\ \bibinfo {author} {\bibfnamefont {I.}~\bibnamefont
  {Wanless}},\ }\href {http://dx.doi.org/10.1007/s00026-005-0261-7} {\bibfield
  {journal} {\bibinfo  {journal} {Annals of Combinatorics}\ }\textbf {\bibinfo
  {volume} {9}},\ \bibinfo {pages} {335} (\bibinfo {year} {2005})},\ \bibinfo
  {note} {10.1007/s00026-005-0261-7}\BibitemShut {NoStop}%
\bibitem [{\citenamefont {Tucker}(2004)}]{Tucker2004}%
  \BibitemOpen
  \bibfield  {author} {\bibinfo {author} {\bibfnamefont {A.}~\bibnamefont
  {Tucker}},\ }\href@noop {} {\emph {\bibinfo {title} {Applied
  Combinatorics}}},\ \bibinfo {edition} {4th}\ ed.\ (\bibinfo  {publisher}
  {John Wiley \& Sons},\ \bibinfo {year} {2004})\BibitemShut {NoStop}%
\bibitem [{\citenamefont {Auluck}(1953)}]{Auluck1953}%
  \BibitemOpen
  \bibfield  {author} {\bibinfo {author} {\bibfnamefont {F.~C.}\ \bibnamefont
  {Auluck}},\ }\href {\doibase doi:10.1017/S0305004100028061} {\bibfield
  {journal} {\bibinfo  {journal} {Mathematical Proceedings of the Cambridge
  Philosophical Society}\ }\textbf {\bibinfo {volume} {49}},\ \bibinfo {pages}
  {72} (\bibinfo {year} {1953})}\BibitemShut {NoStop}%
\bibitem [{\citenamefont {Liu}(1968)}]{Liu1968}%
  \BibitemOpen
  \bibfield  {author} {\bibinfo {author} {\bibfnamefont {C.~L.}\ \bibnamefont
  {Liu}},\ }\href@noop {} {\emph {\bibinfo {title} {Introduction to
  Combinatorial Mathematics}}}\ (\bibinfo  {publisher} {McGraw-Hill},\ \bibinfo
  {year} {1968})\BibitemShut {NoStop}%
\end{thebibliography}%

\end{document}